\documentclass[]{aa} 
\usepackage{graphicx,txfonts,amssymb,natbib}
\renewcommand{\d}{\mathrm{d}} 
\authorrunning{M. Meneghetti et al.}
\titlerunning{Arc sensitivity to cluster ellipticity, asymmetries and
  substructures} 

\begin{document}

\title{Arc sensitivity to cluster ellipticity, asymmetries and  substructures}

 \author{Massimo Meneghetti\inst{1,2}, Rodolfo Argazzi\inst{3}, Francesco
  Pace\inst{1}, Lauro Moscardini\inst{4,8}, Klaus Dolag\inst{5}, Matthias
  Bartelmann\inst{1}, Guoliang Li\inst{6}, Masamune Oguri\inst{7}} 

\institute {$^1$ ITA, Zentrum f\"ur Astronomie, Universit\"at
  Heidelberg, Albert \"Uberle Str. 2, 69120 Heidelberg, Germany \\ $^2$ INAF-Osservatorio
  Astronomico di Bologna, Via Ranzani 1, 40127 Bologna, Italy \\ $^3$
  Dipartimento di Fisica, Universit\`a di Bologna, Via Berti-Pichat 6/2, 40127
  Bologna, Italy \\ $^4$ Dipartimento di Astronomia, Universit\`a di Bologna,
  Via Ranzani 1, 40127 Bologna, Italy \\ $^5$ Max-Planck-Institut f\"ur
  Astrophysik, Karl-Schwarzschild-Str. 1 85748, Garching bei Muenchen,
  Germany\\ $^6$ Shanghai Astronomical Observatory: the partner group of MPA,
  Nandan Road 80, Shanghai 200030, China \\ $^7$ Department of Astrophysical
  Sciences, Princeton University, Peyton Hall, Ivy Lane, Princeton, NJ 08544,
  USA \\ $^8$ INFN, Sezione di Bologna, viale Berti Pichat 6/2, 40127 Bologna, Italy}

\date{\emph{Astronomy \& Astrophysics, submitted}}

\abstract {} {We investigate how ellipticity, asymmetries and substructures
separately affect the ability of galaxy clusters to produce strong lensing
events, i.e. gravitational arcs, and how they influence the arc morphologies
and fluxes. This is important for those studies aiming, for example, at
constraining cosmological parameters from statistical lensing, or at
determining the inner structure of galaxy clusters through gravitational
arcs.}  
{We do so by creating two-dimensional smoothed, differently elliptical
and asymmetric versions of some numerical models. By subtracting these
smoothed mass distributions from the corresponding numerical maps and by
gradually smoothing the residuals before re-adding them to the clusters, we
are able to see how the lensing properties of the clusters react to even small
modification of the cluster morphology. We study in particular by how much
ellipticity, asymmetries and substructures contribute to the strong lensing
cross sections of clusters. We also investigate how cluster substructures
affect the morphological properties of gravitational arcs, their positions and
fluxes.}  
{On average, we find that the contributions of ellipticity,
asymmetries and substructures amount to $\sim 40\%$, $\sim 10\%$ and $\sim
30\%$ of the total strong lensing cross section, respectively. However, our
analysis shows that substructures play a more important role in less
elliptical and asymmetric clusters, even if located at large distances from
the cluster centres ($\sim 1h^{-1}$Mpc). Conversely, their effect is less
important in highly asymmetric lenses. The morphology, position and flux of
individual arcs are strongly affected by the presence of substructures in the
clusters. Removing substructures on spatial scales $\lesssim 50h^{-1}$kpc,
roughly corresponding to mass scales $\lesssim 5 \times
10^{10}h^{-1}\,M_\odot$, alters the image multiplicity of $\sim 35\%$ of the
sources used in the simulations and causes position shifts larger than $5''$
for $\sim 40\%$ of the arcs longer than $5''$.}  
{We conclude that any model
for cluster lens cannot neglect the effects of ellipticity, asymmetries and
substructures. On the other hand, the high sensitivity of gravitational arcs
to deviations from regular, smooth and symmetric mass distributions suggests
that strong gravitational lensing is potentially a powerful tool to measure
the level of substructures and asymmetries in clusters.}


\maketitle

\section{Introduction}
Thanks to the improvements in the quality and in the depth of astronomical
observations, in particular from space, an increasing number of
gravitational arcs has recently been discovered near the centres of many
galaxy clusters \citep[see e.g.][]{BR05.1}. Since the appearance of these
images reflects the shape of the gravitational potential which is responsible
for their large distortions, strong lensing is, in principle, a very powerful
tool for investigating how the matter, in particular the dark component, is
distributed in the inner regions of cluster lenses.

Determining the inner structure of galaxy clusters is one of the major goals
in cosmology, because it should allow us to set important constraints on the
growth of the cosmic structures in the Universe. Moreover, constraining the
mass distribution in the centre of dark matter halos has become increasingly
important in the recent years, since observations of the dynamics of stars in
galaxy-sized systems revealed the presence of a potential problem within the
Cold-Dark-Matter (CDM) scenario. While numerical simulations in this
cosmological framework predict that dark matter halos in a large range of
masses should develop density profiles characterised by an inner cusp,
observations of the rotation curves of dwarf and low-surface-brightness
galaxies suggest that these objects rather have flat density profiles \citep{FL94.1,MO94.1,BU95.1,BU97.1,MG98.1,DA00.1,FI01.1}.

While the centres of galaxies are dominated by stars, which renders it
extremely complicated to derive constraints on the distribution of their dark
matter, galaxy clusters are an alternative and, in many respects, preferable
class of objects for testing the predictions of the CDM model. In fact,
several authors already tried to investigate the inner structure of these
large systems, using and often combining several kinds of observations. Apart
from lensing, the gravitational potential of galaxy clusters can be traced
with several other methods, for example through the emission in the X-ray band
by the hot intra-cluster gas. However, while gravitational lensing directly
probes the matter content of these objects, the other techniques usually rely
on some strong assumptions about their dynamical state and the interaction
between their baryonic and dark matter. For example, it must be often
assumed that the gas is in hydrostatic equilibrium within the dark matter
potential well and that the system is spherically symmetric.

Some ambiguous results were found when comparing the constraints on the inner
structure of clusters as obtained from X-ray and lensing observations. First,
masses estimated from strong lensing are usually larger by a factor of 2-3
than the masses obtained from X-ray observations
\citep{CH03.1,OT04.1}. Deviations from axial symmetry and substructures are
known to be important factors in strong lensing mass estimates \citep[see
e.g.][]{BA95.2,BA96.2,ME03.1,OG05.1,GA05.1}. Second, the constraints on the
inner slope of the density profiles seem to be compatible with a wide range of
inner slopes \citep{ET02.1,LE03.1,AR02.1,SA03.1,BA04.1,GA05.1}.

Apart from the above-mentioned uncertainties affecting the X-ray measurements,
strong lensing observations also have several potential weaknesses. First of
all, arcs are relatively rare events. Frequently, all the constraints which
can be set on the inner structure of clusters via strong lensing depend on a
single or on a small number of arcs and arclets observed near the cluster
core. Second, arcs are the result of highly non-linear effects. This implies
that their occurrence and their morphological properties are very sensitive to
ellipticity, asymmetries and substructures of the cluster matter distribution.

Reversing the problem, this means that, in order to reliably describe the
strong lensing properties of galaxy clusters, all of these effects must be
taken into account. Fitting the positions and the morphology of gravitational
arcs to derive the underlying mass distributions of the lensing clusters,
usually requires to build models with multiple mass components, each of which
is characterised by its ellipticity and orientation \citep[see
e.g.][]{KN93.1,CO05.1,BR05.1}. Even describing the cluster lens population in
a statistical way requires to use realistic cluster models
\citep{ME00.1,ME03.1,ME03.2,OG02.1,OG03.1,DA04.1,HE05.1}. 

Despite the fact that the importance of ellipticity, asymmetries and
substructures for strong lensing appears clearly in many previous studies, many
questions still remain. For example, what is the typical scale of
substructures which contribute significantly to the strong lensing ability of
a cluster?  Where are they located within the clusters? What is the relative
importance of asymmetries compared to ellipticity? Moreover, how do
substructures influence the appearance of giant arcs? All of these open
problems are important for those studies aiming at constraining cosmological
parameters from statistical lensing, or at determining the inner structure of
galaxy clusters through gravitational arcs.

This paper aims at answering to these questions. To do so, we quantify the
impact of ellipticity, asymmetries and substructures by creating differently
smoothed models of the projected mass distributions of some numerical
clusters. We gradually move from one smoothed model to another through a
sequence of intermediate steps.

The plan of the paper is as follows. In Sect.~\ref{sect:nummod}, we discuss the
characteristics of the numerically simulated clusters that we use in this
study; in Sect.~\ref{sect:raytr}, we explain how ray-tracing simulations are
carried out; Sect.~\ref{sect:smooth} illustrates how we obtain smoothed
versions of the numerical clusters; in Sect.~\ref{sect:power}, we suggest a
method to quantify the amount of substructures, asymmetry and ellipticity of
the cluster lenses, based on multipole expansions of their surface density
fields; Sect.~\ref{sect:resu} is dedicated to the discussion of the results of
our analysis. Finally, we summarise our conclusions in Sect.\ref{sect:conclu}.

\section{Numerical models}
\label{sect:nummod}

The cluster sample used in this paper is made of five massive dark matter
halos. One of them, labelled $g8_{\rm hr}$, was simulated with very high mass
resolution, but contains only dark-matter. The others, the clusters $g1$,
$g8$, $g51$ and $g72$ have lower mass resolution but are obtained from
hydro-dynamical simulations which also include gas.

The halos we use here are massive objects with masses $8.1\times
10^{14}\:h^{-1}M_\odot$ ($g72$), $8.6\times 10^{14}\:h^{-1}M_\odot$ ($g51$),
$1.4\times 10^{15}\:h^{-1}M_\odot$ ($g1$) and $1.8\times
10^{15}\:h^{-1}M_\odot$ ($g8$ and $g8_{\rm hr}$) at $z=0.3$. We have chosen
this redshift because it is close to where the strong lensing efficiency of
clusters is the largest for sources at $z_{\rm s} \gtrsim 1$
\citep{LI05.1}.

The clusters were extracted from a cosmological simulation with a box-size of
$479\,h^{-1}\,{\rm Mpc}$ of a flat $\Lambda$CDM model with $\Omega_0=0.3$,
$h=0.7$, $\sigma_8=0.9$, and $\Omega_{\rm b}=0.04$ (see
\citealt{YO01.1}). Using the ``Zoomed Initial Conditions'' (ZIC) technique
\citep{TO97.2}, they were re-simulated with higher mass and force resolution
by populating their Lagrangian volumes in the initial domain with more
particles, appropriately adding small-scale power. The initial displacements
are generated using a ``glass'' distribution \citep{WH96.1} for the Lagrangian
particles. The re-simulations were carried out with the Tree-SPH code GADGET-2
\citep{SP01.1,SP05.1}. For the low resolution clusters, the simulations
started with a gravitational softening length fixed at
$\epsilon=30.0\,h^{-1}\,\mathrm{kpc}$ comoving (Plummer-equivalent) and switch
to a physical softening length of $\epsilon=5.0\,h^{-1}\,\mathrm{kpc}$ at
$1+z=6$.

The particle masses are $m_{\rm DM}=1.13\times 10^9\:h^{-1}M_\odot$ and
$m_{\rm GAS}=1.7\times 10^8\:h^{-1}M_\odot$ for the dark matter and gas
particles, respectively.  For the high-resolution cluster $g8_{\rm hr}$ the
particle mass is $m_{\rm DM}=2.0\times 10^8\:h^{-1}M_\odot$ and the softening
was set to half of the value used for the low resolution runs. Its virial
region at $z=0.3$ contains more than nine million particles, which allow us to
well resolve substructures on scales down to those of galaxies.  Despite the
lower mass resolution with respect to $g8_{\rm hr}$, the other low resolution
clusters also contain several million particles within their virial
radii.

To introduce gas into the high-resolution regions of the low-resolution
clusters, each particle in a control run containing only dark matter was
split into a gas and a dark matter particle. These were displaced by half
the original mean inter-particle distance, so that the centre-of-mass and
the momentum were conserved.

Selection of the initial region was done with an iterative process
involving several low-resolution, dissipation-less re-simulations to
optimise the simulated volume. The iterative cleaning process ensures
that all these haloes are free of contaminating boundary effects up
to at least 3 to 5 times the virial radius. 

The simulations including gas particles follow only the non radiative
evolution of the intra-cluster medium. More sophisticated versions of these
clusters, where radiative cooling, heating by a UV background, and a treatment
of the star formation and feedback processes were included exist and
their lensing properties have been studied in detail by \cite{PU05.1}.

The cluster $g8_{\rm hr}$ is in principle a higher-resolution, dark-matter
only version of the cluster $g8$, which was simulated with non-radiative gas
physics. Nevertheless the two objects can only be compared
statistically. Indeed, the introduction of the gas component as well as the
increment of the mass resolution introduce small perturbations to the initial
conditions, which lead to slightly different time evolutions of the simulated
halos. Furthermore, also the presence of gas and its drag due to pressure lead
to significant changes in the assembly of the halo. A detailed discussion of
such differences can be found in \cite{PU05.1}. For this reason, the lensing
properties of $g8$ and $g8_{hr}$ at $z=0.3$ are not directly comparable.

\section{Lensing simulations}
\label{sect:raytr}

Ray-tracing simulations are carried out using the technique described in
detail in several earlier papers (e.g.~\citealt{BA98.2,ME00.1}).

We select a cube of $6\,h^{-1}$Mpc comoving side
length, centred on the halo centre and containing the high-density
region of the cluster. The particles in this cube are used for
producing a three-dimensional density field, by interpolating their
position on a grid of $1024^3$ cells using the {\em Triangular Shaped
Cloud} method \citep{HO88.1}. Then, we project the three-dimensional
density field along the coordinate axes, obtaining three surface
density maps $\Sigma_{i,j}$, used as lens planes in the following
lensing simulations. 

The lensing simulations are performed by tracing a bundle of $2048
\times 2048$ light rays through a regular grid, covering the central
sixteenth of the lens plane. This choice is driven by the necessity to
study in detail the central region of the clusters, where critical
curves form, taking into account the contribution from the surrounding
mass distribution to the deflection angle of each ray.

Deflection angles on the ray grid are computed using the method
described in \citet{ME00.1}. We first define a grid of $256\times256$
``test'' rays, for each of which the deflection angle is calculated by
directly summing the contributions from all cells on the surface
density map $\Sigma_{i,j}$,
\begin{equation}
  \vec \alpha_{h,k}=\frac{4G}{c^2}\sum_{i,j} \Sigma_{i,j} A
  \frac{\vec x_{h,k}-\vec x_{i,j}}{|\vec x_{h,k}-\vec x_{i,j}|^2}\;,
\end{equation}  
where $A$ is the area of one pixel on the surface density map and
$\vec x_{h,k}$ and $\vec x_{i,j}$ are the positions on the lens plane
of the ``test'' ray ($h,k$) and of the surface density element
($i,j$). Following \cite{WA98.2}, we avoid the divergence when the
distance between a light ray and the density grid-point is zero by
shifting the ``test'' ray grid by half-cells in both directions with
respect to the grid on which the surface density is given. We then
determine the deflection angle of each of the $2048\times2048$ light
rays by bi-cubic interpolation between the four nearest test rays.

The position $\vec y$ of each ray on the source plane is calculated by
applying the lens equation. If $\vec y$ and $\vec x$ are the angular
positions of source and image from an arbitrarily defined optical axis
passing through the observer and perpendicular to the lens and source
planes, this is written as
\begin{equation}
  \vec y = \vec x -\frac{D_{\rm ls}}{D_{\rm s}}\vec \alpha(\vec x)\;,
\end{equation}
where $D_{\rm ls}$ and $D_{\rm s}$ are the angular diameter distances
between the lens and the source planes and between the observer and
the source plane, respectively.

Then, a large number of sources is distributed on the source plane. We
place this plane at redshift $z_\mathrm{s}=2$. Keeping all sources at
the same redshift is an approximation justified for the purposes of
the present case study, but the recent detections of arcs in
high-redshift clusters \citep{ZA03.1,GL03.1} indicate that more
realistic simulations will have to account for a wide source redshift
distribution.

The sources are elliptical with axis ratios randomly drawn from
$[0.5,1]$. Their equivalent diameter (the diameter of the circle
enclosing the same area of the source) is $r_\mathrm{e}=1''$. They
are distributed on a region on the source plane corresponding to one
quarter of the field of view where rays are traced. As in our earlier
studies, we adopt an adaptive refinement technique when placing
sources on their plane. We first start with a coarse distribution of
$32\times32$ sources and then increase the source number density
towards the high-magnification regions of the source plane by adding
sources on sub-grids whose resolution is increased towards the lens
caustics. This increases the probability of producing long arcs and
thus the numerical efficiency of the method. In order to compensate
for this artificial source-density enhancement, we assign a
statistical weight to each image for the following statistical
analysis which is proportional to the area of the sub-grid cell on
which the source was placed.

By collecting rays whose positions on the source plane lie within any
single source, we reconstruct the images of background galaxies and
measure their length and width. Our technique for image detection and
classification was described in detail by \cite{BA94.1} and used by
\cite{ME00.1,ME01.1,ME03.2,ME03.1,TO04.1} and \cite{ME05.1}. 
The modifications recently suggested by \cite{PU05.1} for increasing the
accuracy of the measurements of the arc properties have been included in our
code. The simulation process ends in a catalogue of images which is
subsequently analysed.

\section{Smoothed representations of the cluster}
\label{sect:smooth}
We aim here at separating the effects of substructures, ellipticity and
asymmetries on the strong lensing efficiency of our numerical clusters.
We work in two dimensions, starting from the
projected two-dimensional mass distribution of each lens. Since the lens'
surface density profile plays a crucial role for strong lensing, we keep it
fixed and vary only the shape of the density contours.

As a first step, we construct a fiducial model for the smooth mass distribution
of the lens. This is done by measuring the ellipticity and the position angle
of the surface density contours as function of radius in the projected mass
map. The projected density is measured in circles of increasing radii $x$.
We determine the quadrupole moments of the density distribution in each
aperture,

\begin{equation}
  S_{ij}(x) = \frac{\int \d^2 x \Sigma(\vec x)
  (x_i-x_{c,i})(x_j-x_{c,j})}{\int \d^2 x \Sigma(\vec x)} \, , \, i,j \in
  (1,2) ,    
\end{equation}

where $\vec x_c$ denotes the position of the cluster centre and the integrals
are extended to the area enclosed by the aperture. Both the ellipticity
$\epsilon$ and the position angle $\phi$ of the iso-density contour
corresponding to the chosen aperture radius are derived from the elements of
the tensor $S_{ij}(r)$:
\begin{eqnarray}
  \epsilon(x) & =
  &\sqrt{\frac{(S_{11}-S_{22})^2+4S_{12}^2}{(S_{11}+S_{22})^2}} \, , \\
  \phi(x) & = & \frac{1}{2} \arctan{\frac{2S_{12}}{S_{11}-S_{22}}} \, .
\end{eqnarray}

\begin{figure*}[ht]
  \includegraphics[width=\hsize]{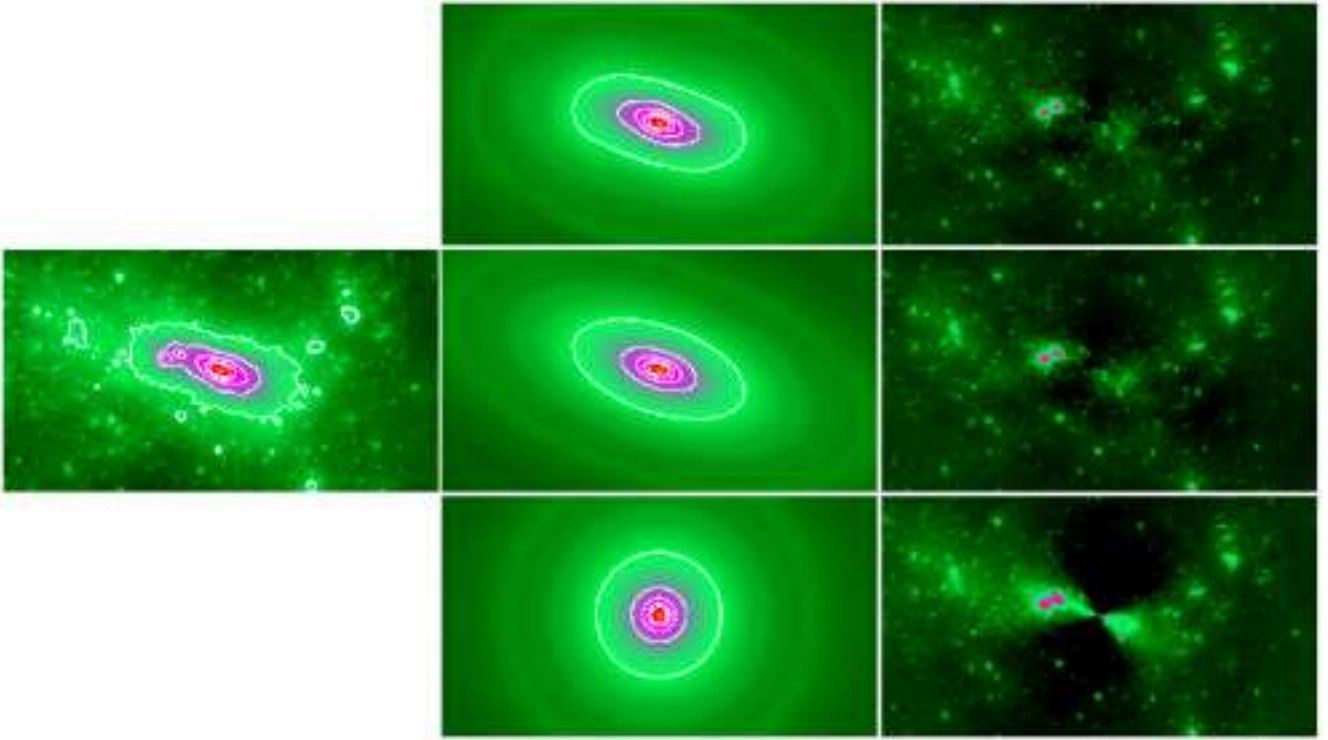}
\caption{Different smoothing sequences of the projection along
  the $x$-axis of cluster $g8_{\rm hr}$. The original map is shown on the
  left. Top panels: smoothing with variable ellipticity and orientation of the
  iso-density contours; middle panels: smoothing assuming fixed ellipticity
  and position angle (see text for more details); bottom panels: smoothing
  assuming axial symmetry. From left to right, each row of panels show 
  the smoothed map and the residuals, obtained by subtracting the smoothed 
  from the original map. The contour levels in the density maps start at $\sim
  3.6\times10^{15}\,h\,M_\odot\,$Mpc$^{-2}$ and are spaced by
  $5\times10^{15}\,h\,M_\odot\,$Mpc$^{-2}$. The horizontal side length of each
  panel is $6\,h^{-1}$Mpc.}
\label{fig:smooth}
\end{figure*} 

The ellipticity and position angle profiles are used in combination with the
mean surface density profile $\overline{\Sigma}(x)$ of the lens for
constructing a smoothed surface density map,
\begin{equation}
  \Sigma^{\rm sm}(\vec x)=\overline{\Sigma}(x_{\epsilon,\phi})\;,
  \label{eq:smooths}
\end{equation}
where the equivalent radius $x_{\epsilon,\phi}$ is given by
\begin{eqnarray}
  x_{\epsilon,\phi} & = &
  \left(\frac{[x_1\cos\phi(x)+x_2\sin\phi(x)]^2}{[1-e(x)]} \right. \nonumber
  \\ 
  &+&  \left. [-x_1\sin\phi(x)+x_2\cos\phi(x)]^2[1+e(x)]\right)^{1/2}\;,  
  \label{eq:eqrad}
\end{eqnarray}
and
$e=1-\sqrt{1-\epsilon^2}/(1+\epsilon)=1-\sqrt{1-\epsilon}/\sqrt{1+\epsilon}
\approx \epsilon$ (for $\epsilon \ll 1$).

The resulting map conserves the mean surface density profile of the cluster
and reproduces well the twist of its iso-density contours, i.e. the asymmetries
of the projected mass distribution. This is shown in
Fig.~\ref{fig:smooth}. The panel on the left column shows the original surface
density map for one projection of the high-resolution cluster $g8_{\rm hr}$.
Contour levels start at $\sim 3.6\times10^{15}\,h\,M_\odot\,$Mpc$^{-2}$ and
are spaced by $5\times10^{15}\,h\,M_\odot\,$Mpc$^{-2}$. The top-middle panel
shows the smoothed map obtained from Eq.~(\ref{eq:smooths}). In the rest
of the paper, we will call this smoothed model the ``asymmetric''
model. The same colour scale and spacing of the contour levels as in the first
panel are used.  In the smoothed map, the substructures on all scales are
removed and redistributed in elliptical shells around the cluster.  Comparing
the strong lensing properties of the original and of the smoothed map, we can
therefore quantify {\em the net effect of substructures} on the cluster strong
lensing efficiency. By subtracting the smoothed from the original map, we
obtain a residual map showing which substructures will not contribute to
lensing after smoothing. We plot this residual map in the right panel in the
upper row of Fig.~\ref{fig:smooth}.

Similarly, the effects of other cluster properties can be separated. For
example, we can remove asymmetries and deviations from a purely elliptical
projected mass distribution by disabling the twist of the iso-density
contours in our smoothing procedure. For doing so, we fix the ellipticity and
the position angle to a constant value, $\epsilon=\epsilon_{\rm crit}$ and
$\phi=\phi_{\rm crit}$. We choose $\epsilon_{\rm crit}$ and $\phi_{\rm crit}$
to be those measured in the smallest aperture containing the cluster critical
curves. A smoothed map of the cluster is created as explained earlier. The
results are shown in the middle panels of Fig.~\ref{fig:smooth}. A comparison
of the lensing properties of this new ``elliptical'' model with those of the
previously smoothed map allows us to quantify {\em the effect of asymmetries},
which are large-scale deviations from elliptical two-dimensional mass
distributions.

Finally, even the cluster ellipticity can be removed, still preserving the same
mean density profile. This is easily done by inserting $\epsilon=0$ in
Eq.~(\ref{eq:smooths}). The resulting smoothed map and the residuals obtained
by subtracting it from the original projected mass map are shown in the bottom
panels of Fig.\ref{fig:smooth}. If we compare the lensing efficiency of such
an ``axially symmetric'' model to that of the previously defined elliptical
model, we quantify {\em the effect of ellipticity} on the cluster strong
lensing properties.

For each smoothing method, we simulate lensing of background galaxies not only
for the extreme cases of the totally smoothed maps but also for partially
smoothed mass distributions. Adding the residuals $R$ to the smoothed map, the
original surface density map of the cluster is indeed re-obtained,
\begin{equation}
  \Sigma(\vec x)=\Sigma^{\rm sm}(\vec x)+R(\vec x) \,.
\end{equation}
Substructures of different sizes can be gradually removed from the cluster
mass distribution by filtering the residual map with some filter function of
varying width before re-adding it to the totally smoothed map. This can be
done, for example, by convolving the residual map $R$ with a Gaussian $G$ of
width $\sigma$:
\begin{eqnarray}
  \widetilde{R}(\vec x,\sigma)&=&R(\vec x)*G(\vec x,\sigma) \, ,  \\
  G(\vec x,\sigma)&=&\frac{1}{2\pi \sigma^2}
  \exp{\left(-\frac{\vec{x}^2}{2\sigma^2}\right)} \, .
\end{eqnarray}
The width $\sigma$ defines the scale of the substructures which
will be filtered out of the mass map. Surface density maps with
intermediate levels of substructures are finally obtained by adding the
filtered residuals to the smoothed map,
\begin{equation}
  \widetilde{\Sigma}(\vec x,\sigma)=\Sigma^{\rm sm}(\vec x)+\widetilde{R}(\vec
  x,\sigma) \,.
\end{equation}
 
\begin{figure}[t!]
  \includegraphics[width=\hsize]{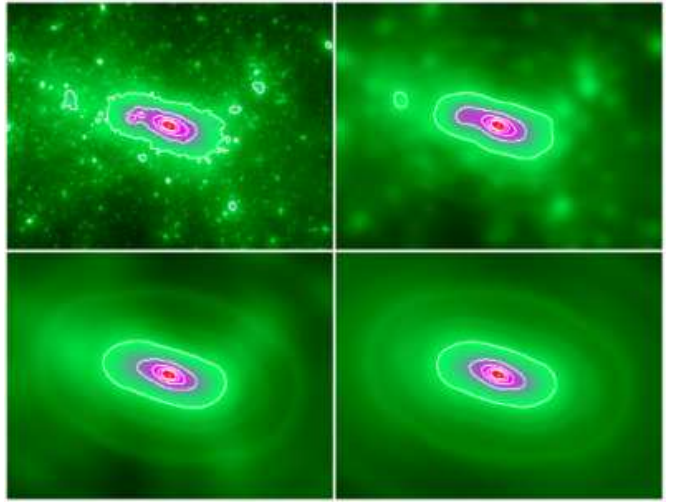}
\caption{Surface density maps of the same cluster projection as shown
  in Fig.~\ref{fig:smooth}, smoothed with increasing smoothing length from
  the top left to the bottom right panels. The background smoothed model is
  the one shown in the upper right panel of Fig.~\ref{fig:smooth}. The
  smoothing lengths in the four panels are $0,47,141$ and $470\,h^{-1}$kpc
  comoving, respectively. The horizontal side length of each panel is
  $6\,h^{-1}$Mpc comoving.}
\label{fig:incsmooth}
\end{figure}

This procedure allows us to investigate what is {\em the characteristic scale
  of substructures which are important for strong lensing}.  Moreover,
comparing how the lensing properties of different clusters react to smoothing, 
we can quantify {\em the impact of substructures in halos with
  different degrees of asymmetry and ellipticity}.  A sequence of smoothed
versions of the same cluster model is shown in Fig~\ref{fig:incsmooth}. The
smoothing length $\sigma$ is $0,47,141$ and $470\,h^{-1}$ comoving kpc from the
top left to the bottom right panel, respectively.

Another important issue is to understand {\em where the substructures must be
  located in order to have a significant impact} on the strong lensing
  properties of clusters. To address this problem, we remove from the clusters
  the substructures located outside apertures of decreasing equivalent radius
  $x_{\epsilon,\phi}$. Again, this is done by modifying the residuals of
  the smoothed maps. We multiply the residual map with the function,
\begin{equation}
  T(\vec x_{\epsilon,\phi},l)=\left\{
    \begin{array}{r@{\quad \quad}l}
      1 & (x_{\epsilon,\phi}<l) \\
      \exp{\left(-\frac{\vec x_{\epsilon,\phi}^2}{2\sigma_{\rm cut}^2}\right)}
      & (x_{\epsilon,\phi} \ge l) 
    \end{array}\right.\;,
  \label{eq:wf}
\end{equation}
where $\sigma_{\rm cut}=100\,h^{-1}$kpc comoving and $l$ is the equivalent
distance beyond which the substructures are suppressed. The Gaussian tail of
the window function was applied to avoid sudden discontinuities in the
surface density maps.

\begin{figure}[t!]
  \includegraphics[width=\hsize]{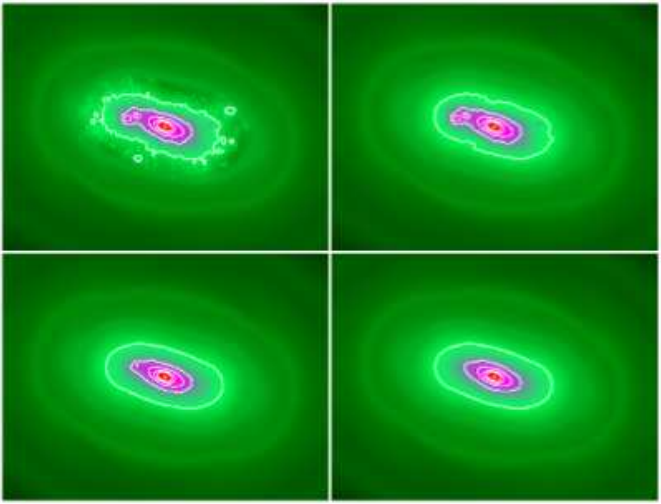}
\caption{Surface density maps of the same cluster projection shown
  in Fig.~\ref{fig:smooth}, suppressing the substructures in shells of
  decreasing equivalent radius from the top left to the bottom right panels.
  The background smoothed model is the one shown in the upper right panel of
  Fig.~\ref{fig:smooth}. The equivalent radii beyond which the substructures
  are removed are $1174,704,352$ and $235\,h^{-1}$kpc comoving, respectively.
  The horizontal side length of each panel is $6\,h^{-1}$Mpc comoving.}
\label{fig:subrad}
\end{figure} 

Results of the removal of cluster substructures at different radii are shown
in Fig.~\ref{fig:subrad}. From the top left to the bottom right panel, the
cut-off scales $l$ are $1174,704,352$ and $235\,h^{-1}$kpc comoving,
respectively. The residual maps filtered with the window function
(\ref{eq:wf}) were re-added to the totally smoothed asymmetric maps for
obtaining surface density distributions with the desired level of
substructures within a given equivalent radius.

\section{Quantifying the amount of substructures, asymmetry and ellipticity}
\label{sect:power}

\begin{figure}[t!]
  \includegraphics[width=\hsize]{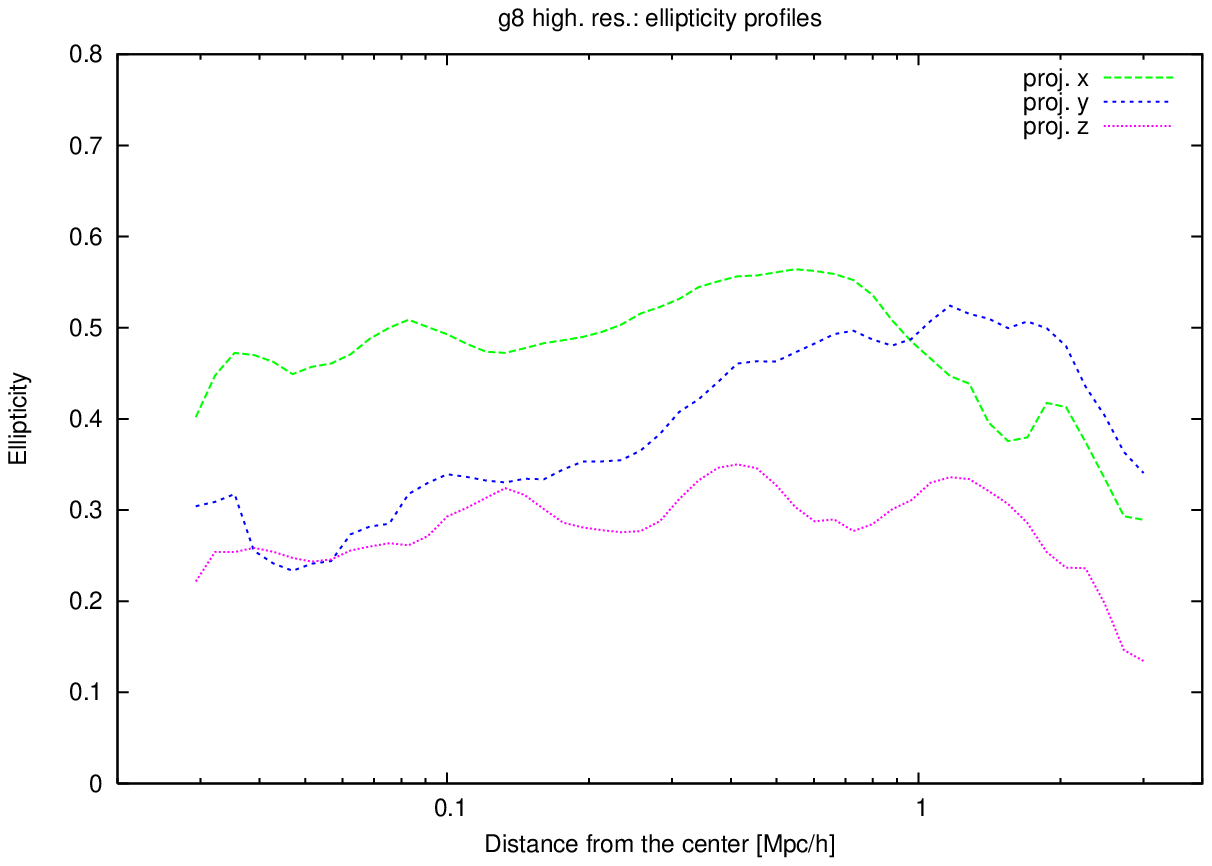}
  \includegraphics[width=\hsize]{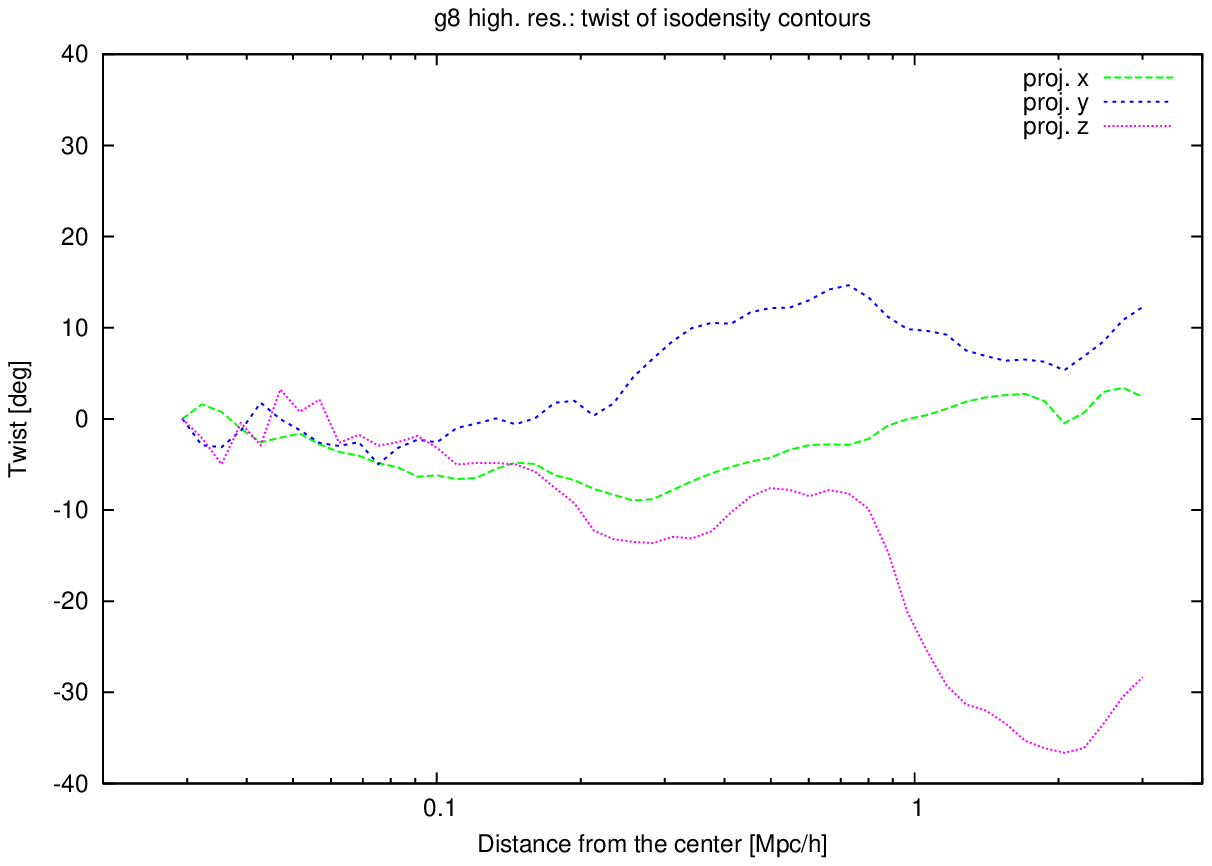}
\caption{Variations of ellipticity (top panel) and position angle (bottom
  panel) of the iso-density contours of the three projections of cluster
  $g8_{\rm hr}$ as function of the distance from the cluster centre. The
  cluster has a virial radius of $\sim 2.7h^{-1}$Mpc.}
\label{fig:ellphi}
\end{figure}

The variations of ellipticity and position angle of the iso-density contours
are given by the functions $\epsilon(x)$ and $\phi(x)$, which were defined in
the previous section. These are shown for the three projections of cluster
$g8_{\rm hr}$ in Fig.~\ref{fig:ellphi}. They illustrate that the projection
along the $x$-axis of this cluster, shown in Fig.~\ref{fig:smooth}, is the
most elliptical at the relevant radii, with an ellipticity which grows from
$\epsilon\sim 0.4$ to $\sim0.58$ within the inner $\sim 1
\,h^{-1}$Mpc. The
iso-density contours have almost constant orientation in this projection. The
projection along the $y$-axis exhibits the largest variations of ellipticity
in the central region of the cluster, with $\epsilon$ growing from $\sim 0.25$
to $\sim 0.52$. It is also characterised by a large twist of the iso-density
contours, whose orientations change by up to $\sim 20$ degrees. When projected
along the $z$-axis, the cluster appears more circular and with fairly constant
ellipticity ranging between $\sim 0.22$ and $\sim 0.32$. The twist of the
iso-density contours is moderate within the inner $1\,h^{-1}$Mpc.

We now quantify the amount of substructure within the numerical
clusters by means of multipole expansions of their surface density maps
\citep{ME03.1}.

Briefly, starting from the particle positions in the
numerical simulations, we compute the surface density $\Sigma$ at
discrete radii $x_n$ and position angles $\theta_k$ taken from
$[0,1.5]\,h^{-1}\,\mbox{Mpc}$ and $[0,2\pi]$, respectively. For any
$x_n$, each discrete sample of data $\Sigma(x_n,\theta_k)$ is expanded
into a Fourier series in the position angle,
\begin{equation}
  \Sigma(x_n,\theta_k)=\sum_{l=0}^\infty\,S_l(x_n)
  \mathrm{e}^{-\mathrm{i}l\theta_k}\;,
\label{eq:expansion}
\end{equation}
where the coefficients $S_l(x_n)$ 
\begin{equation}
  S_l(x_n)=\sum_{k=0}^\infty\,\Sigma(x_n,\theta_k)
  \mathrm{e}^{\mathrm{i}l\theta_k}\;,
\end{equation}
can be computed using fast-Fourier techniques. 

We define the power spectrum $P_l(x_n)$ of the multipole expansion $l$ as
$P_l(x_n)=|S_l(x_n)|^2$. As discussed by \citet{ME03.1}, the amount of
substructure, asymmetry and ellipticity in the mass distributions of the
numerically simulated cluster at any distance $x_n$ from the main clump can be
quantified by defining an integrated power $P_\mathrm{circ}(x_n)$ as the sum of
the power spectra over all multipoles, from which the monopole is
subtracted in order to remove the axially symmetric contribution,
\begin{equation}
  P_\mathrm{circ}(x_n)=\sum_{l=0}^\infty\,P_l(x_n) - P_0(x_n)\;.
\end{equation}
This quantity measures the deviation from a circular distribution of the
surface mass density at a given distance $x_n$ from the cluster centre.

In a fully analogous way, we can quantify the deviation from a purely elliptical
surface mass density by subtracting from $P_\mathrm{circ}$ the quadrupole term
$P_2(x_n)$:
\begin{equation}
  P_\mathrm{ell}(x_n)=P_\mathrm{circ}(x_n) - P_2(x_n)\;.
\end{equation}

Separating the effects of asymmetries and substructures by means of
contributions to the multipoles is not an easy task. The two terms are mixed
together in the multipole expansion. In this paper, asymmetries are assumed to
be large-scale deviations from a purely elliptical mass distribution, which
result in variations of ellipticity and position angle of the iso-density
contours as functions of radius. Their contribution to the azimuthal multipole
expansion is mostly contained in the low-{\em l} tail. On the other hand,
for small $x_n$ even relatively small substructures subtend large angles with
respect to the cluster centre. The dipole term itself contains a contribution
from substructures. Assuming that substructures are localised lumps of matter
whose angular extent is $\lesssim
\pi/2$ and then contribute to all multipoles of order larger than $l\sim 2$,
we quantify the amount of substructures at radius $x_n$ in the cluster by means
of the quantity
\begin{equation}
  P_\mathrm{sub}(x_n)=P_\mathrm{ell}(x_n)-P_1(x_n)\;,
\end{equation}
where the power corresponding to the dipole, $P_1(x_n)$ has been subtracted
from $P_\mathrm{ell}(x_n)$.

\begin{figure}[t]
  \includegraphics[width=\hsize]{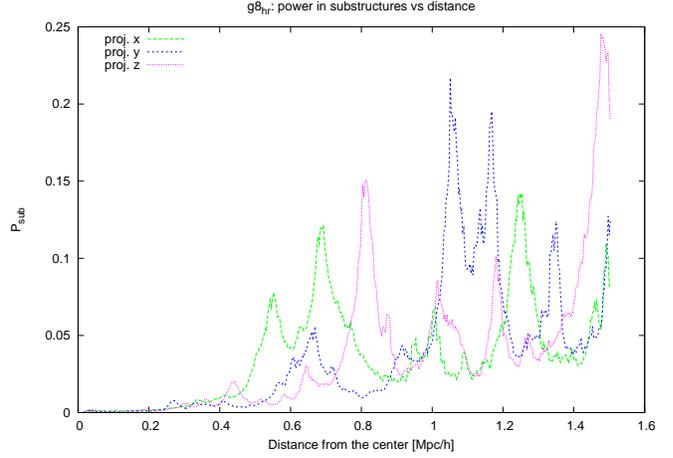}
\caption{Power in substructures as a function of distance from the
  cluster centre for the three projections along the $x$-, $y$- and $z$-axes of
  cluster $g8_{\rm hr}$.}
\label{fig:powers}
\end{figure} 

Fig.\ref{fig:powers} shows the radial profiles of $P_{\rm sub}$ for the three
projections along the $x$, $y$ and $z$-axes of cluster $g8_{\rm hr}$. Peaks
along the curves indicate the presence of substructures. The amplitude of the
peaks is a growing function of the mass of the substructures. Clearly, in the
projection along the $x$-axis massive lumps of matter are located at distances
of $\sim 550$ and $\sim 700\,h^{-1}$kpc from the cluster centre. Substructures
are abundant also at radii of $\sim 1$ and $\sim 1.25\,h^{-1}$Mpc. This is
visible in the left panel of Fig.\ref{fig:smooth}. Instead, the most dominant
substructures in the projection along the $y$-axis are located outside the
region of radius $1\,h^{-1}$Mpc. Only a relatively small peak is observed at
$\sim 650\,h^{-1}$kpc from the centre. Finally, when projected along the
$z$-axis, the cluster contains a large amount of substructures at the distance
of $\sim 800\,h^{-1}$kpc from the centre. Other peaks are located at radii
$>1\,h^{-1}$Mpc.

\section{Results}
\label{sect:resu}

In this section, we describe the lensing properties of the numerical clusters
in the sample we have analysed and quantify the impact of ellipticity,
asymmetries and substructures on their ability to produce arcs.

\subsection{Magnification and caustics}
\label{sect:mag}

The lensing properties of the two-dimensional mass distributions generated
using the previously explained methods can be easily determined using the
standard ray-tracing technique described in Sect.~\ref{sect:raytr}.

The ability of galaxy clusters to produce strong lensing events is expected to
reflect both the presence of substructures embedded into their halos and
the degree of ellipticity and  asymmetry of their mass
distributions. Indeed, all of these factors contribute to increase the shear
field of the clusters. This was shown for example by \cite{TO04.1} and later
confirmed by \cite{ME04.1} and \cite{FE05.1}, who found that the passage of
substructures through the cluster cores can enhance the lensing cross
section for the formation of arcs with large length-to-width ratios by orders
of magnitude. \cite{ME03.1} show that elliptical models with realistic
density profiles produce a number of arcs larger by a factor of ten compared to
axially-symmetric lenses with the same mass. Analogous results were
obtained by \cite{OG03.1}, who compared the lensing efficiency of triaxial and
spherically symmetric halos, and more recently by \cite{HE05.1}.

\begin{figure}[t]
  \includegraphics[width=\hsize]{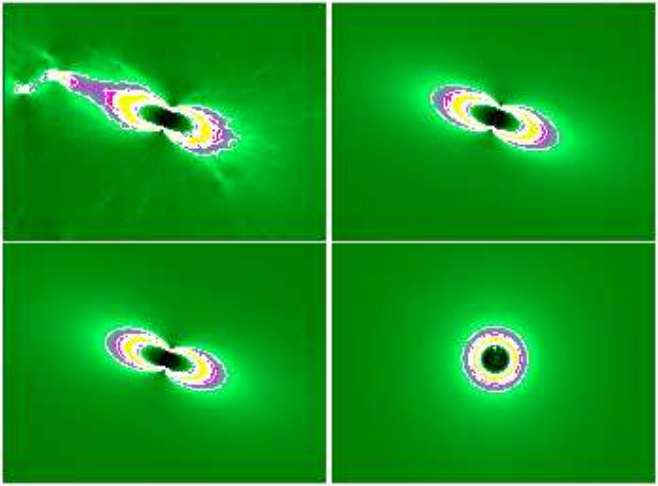}
\caption{Maps of the tangential-to-radial magnification ratio for the same 
cluster projection showed in the previous figures. From the top left to the
  bottom right panel, we show the maps corresponding to the original cluster
  and to three smoothed versions of it: using the asymmetric, the elliptical
  and the axially symmetric background models.}
\label{fig:lw_maps}
\end{figure}

\begin{figure}[h]
  \includegraphics[width=\hsize]{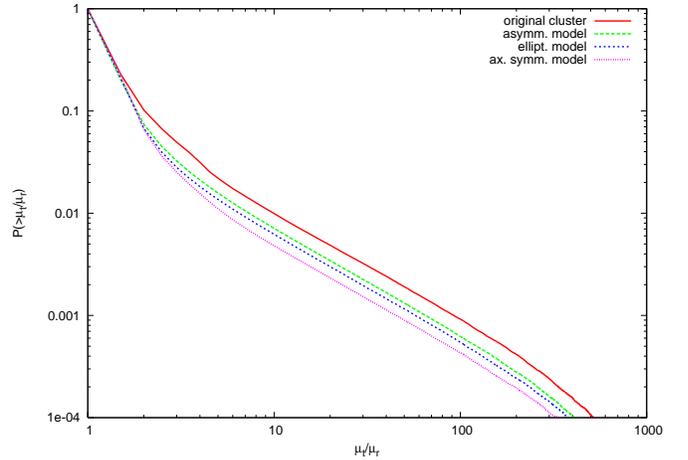}
\caption{Cumulative distributions of tangential-to-radial magnifications in the
maps showed in Fig.\ref{fig:lw_maps}.}
\label{fig:lw_cum}
\end{figure}

\begin{figure*}[t]
  \includegraphics[width=0.33\hsize]{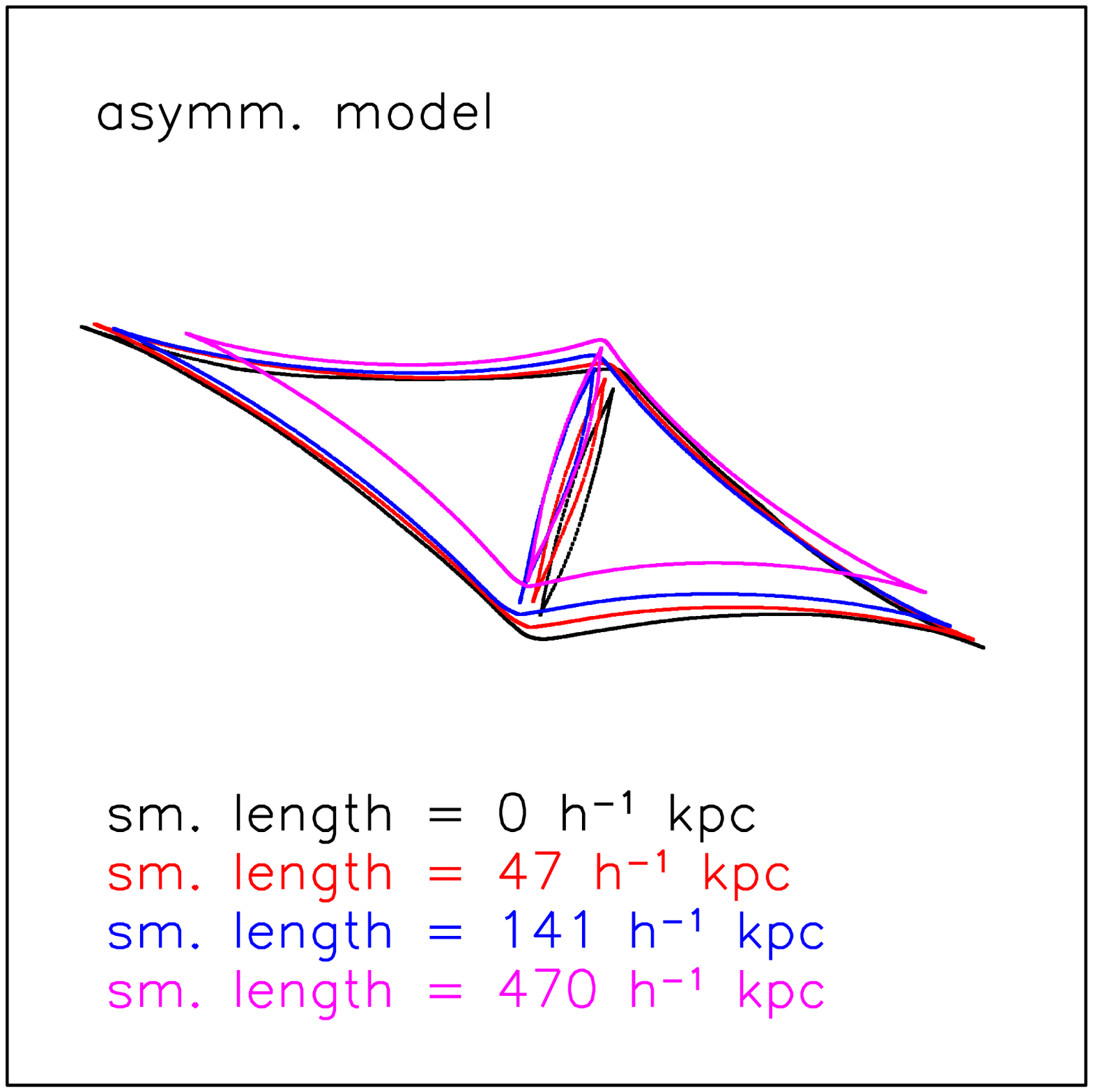}
  \includegraphics[width=0.33\hsize]{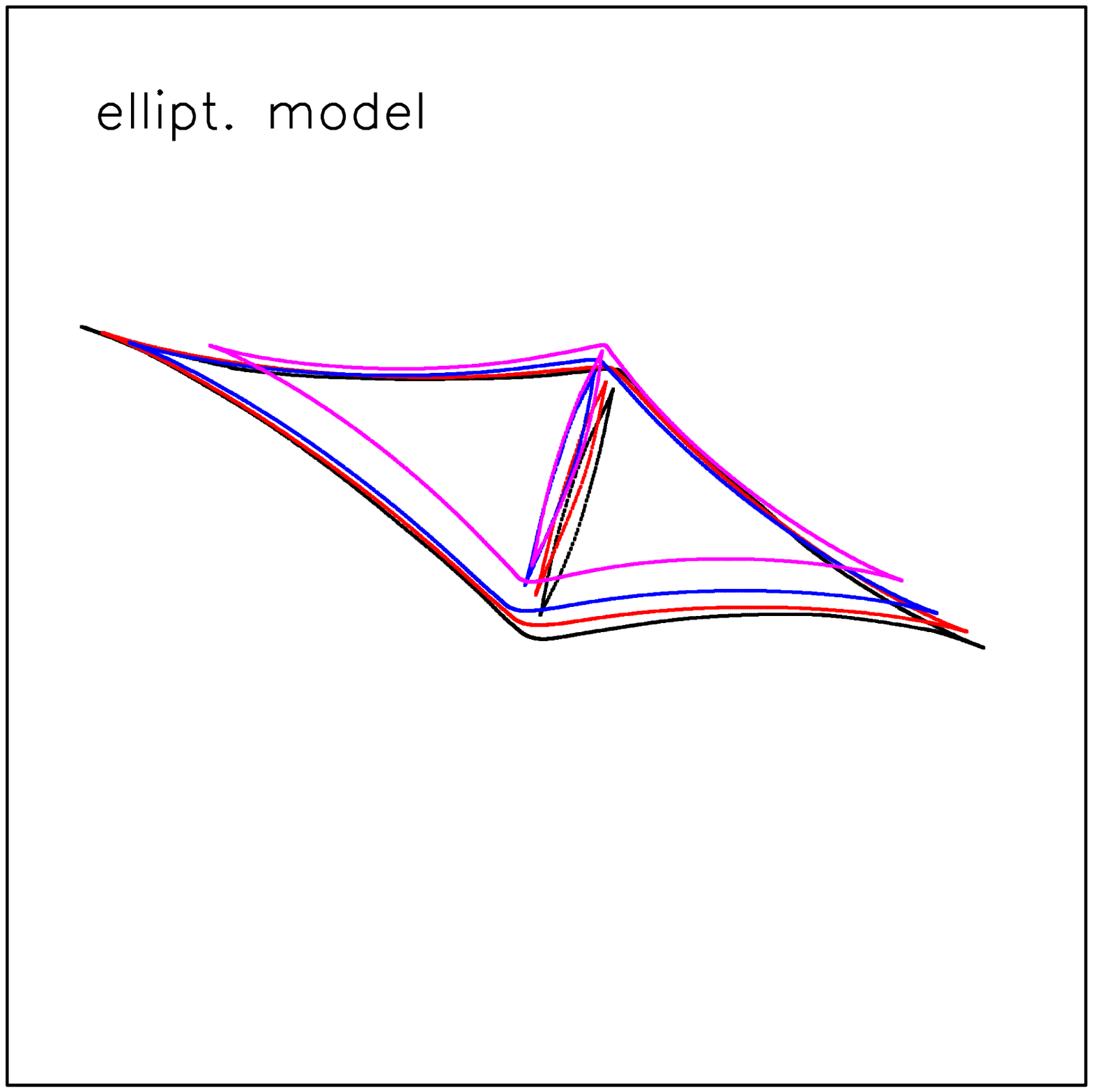}
  \includegraphics[width=0.33\hsize]{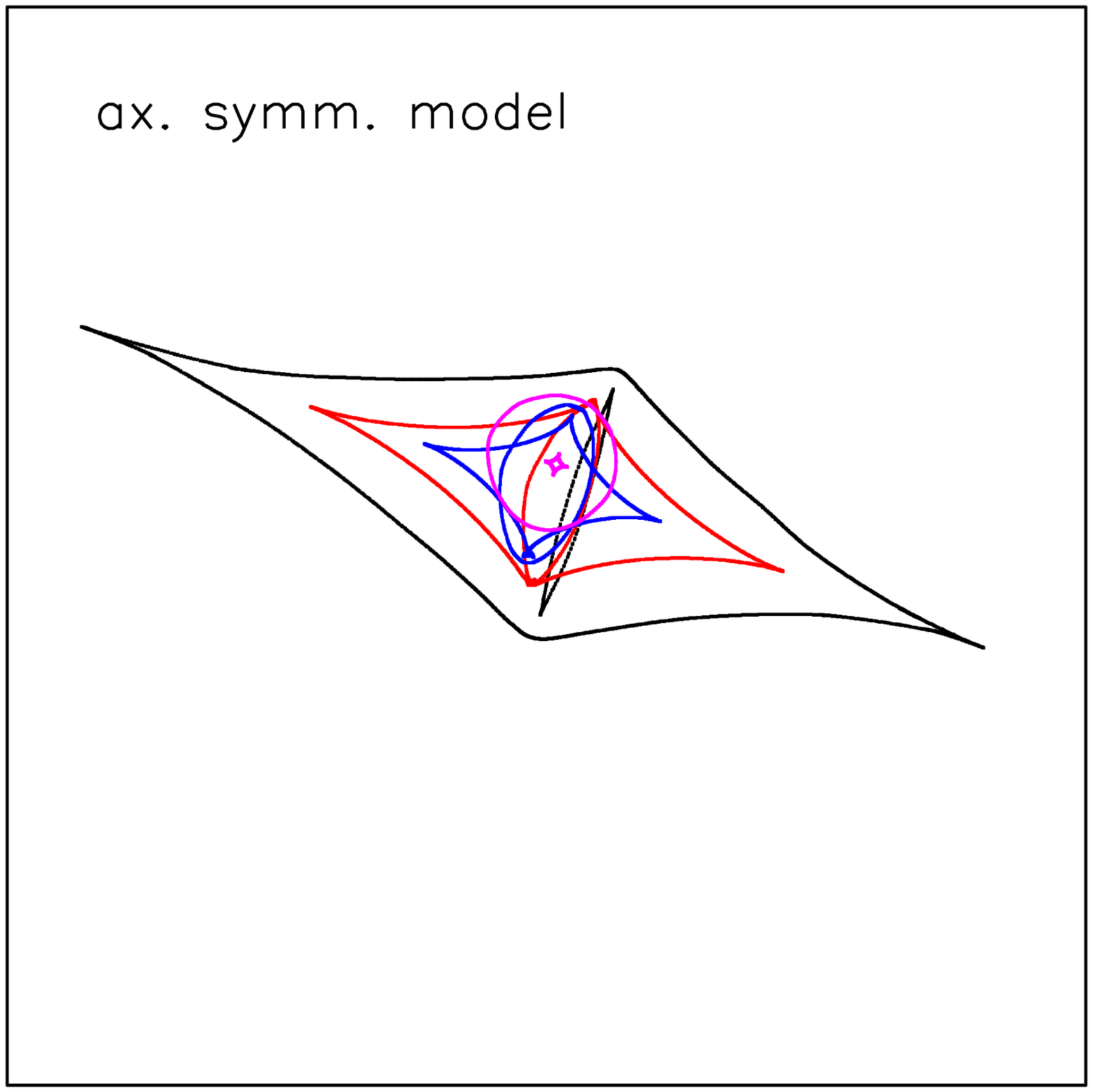}
\caption{Lens caustics of the mass distribution of the cluster projected along
  the $x$-axis as obtained after smoothing using several smoothing lengths and
  assuming different background models: asymmetric model (left panel),
  elliptical model (central panel) and axially symmetric model (right
  panel). The side length of each panel is $500\,h^{-1}$kpc comoving,
  corresponding to $\sim 1'$.}
\label{fig:cau_smooth}
\end{figure*}

By smoothing the two-dimensional mass distribution of the clusters, both the
levels of substructures and asymmetries are decreased. Thus, we expect their
ability to produce highly distorted arcs to be somewhat reduced. This
expectation is supported by the fact that the regions of the lens plane where
the tangential-to-radial magnification ratio exceeds a given threshold shrink
significantly when the smoothing is applied. This is shown in
Fig.~\ref{fig:lw_maps}. The map of the tangential-to-radial magnification
ratio of the original cluster (top right panel) is compared to those obtained
by smoothing its surface density map using the asymmetric (top left panel),
the elliptical (bottom left panel) and the axially symmetric (bottom right
panel) background models. The cumulative distributions of the pixel values in
these maps are displayed in Fig.\ref{fig:lw_cum}. The probability of having
pixels where the tangential-to-radial magnification ratio exceeds the minimal
value decreases at least by a factor of two, due to removal of substructures,
asymmetries and ellipticity. This does not imply that the cross section for
arcs with large length-to-width ratio decreases by the same amount, since the
excess of pixels with large tangential-to-radial magnification ratio in the
unsmoothed map is in part due to isolated lumps of matter whose angular scale
is similar or smaller than the angular scale of the sources.

By definition, the lensing cross section, which measures a cluster's ability
to produce arcs, is an area encompassing the lens' caustics. Thus, the more
extended the caustics are, the larger is generally the lensing cross
section. In Fig.~\ref{fig:cau_smooth} we show how the caustic shape changes as
the smoothing length is varied. Results are shown for each of the smoothing
schemes applied. As expected, the caustics shrink as the smoothing length
increases. Comparable trends are found for the asymmetric and elliptical
background models, for which the change of the caustic length is not
dramatic. On the other hand, if the cluster surface density is gradually
smoothed towards an axially symmetric distribution, the evolution of the lens'
caustics is much stronger.

\begin{figure}[h]
  \includegraphics[width=\hsize]{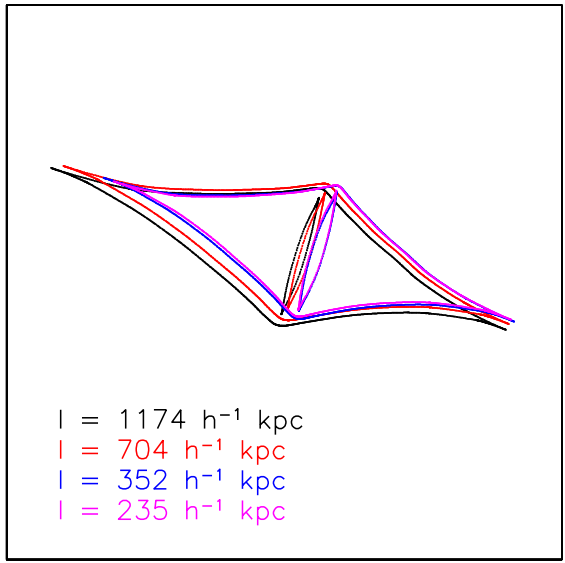}
\caption{Lens' caustics obtained after removing substructures from the region
  outside a given radius, as given by the labels in the figure. The side
  length is $500\,h^{-1}$kpc comoving, corresponding to $\sim 1'$. The figure
  refers to the cluster projection along the $x$ axis.}
\label{fig:cau_sub}
\end{figure}

Similar reductions of the caustic sizes are found when suppressing the
substructures outside a given radius. This is shown in
Fig.~\ref{fig:cau_sub}. Clearly, substructures at
distances of the order of $1\,h^{-1}$Mpc already play a significant role for
strong lensing. Although they are located far away from the cluster critical
region, the external shear they produce is remarkable and determines an
expansion of the lens' caustics. 

\begin{figure*}[t]
  \includegraphics[width=0.33\hsize]{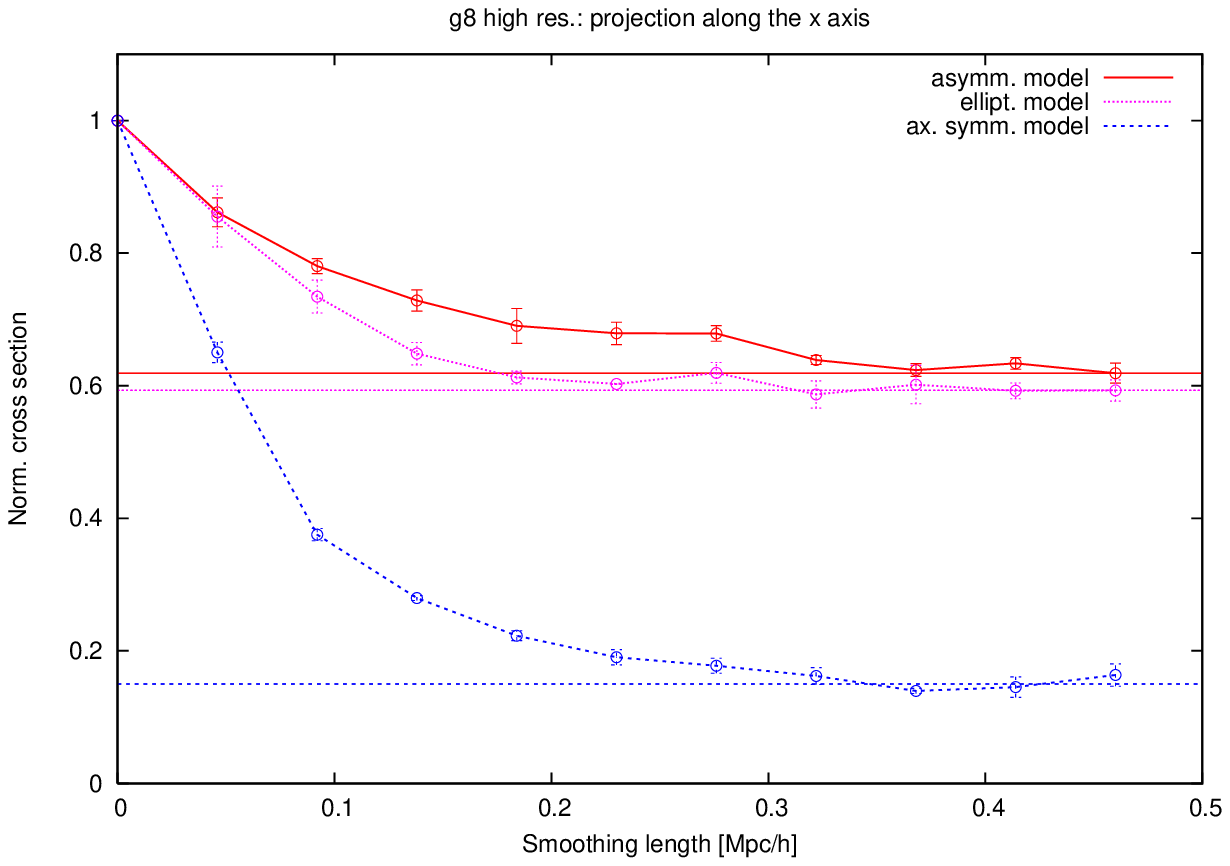}
  \includegraphics[width=0.33\hsize]{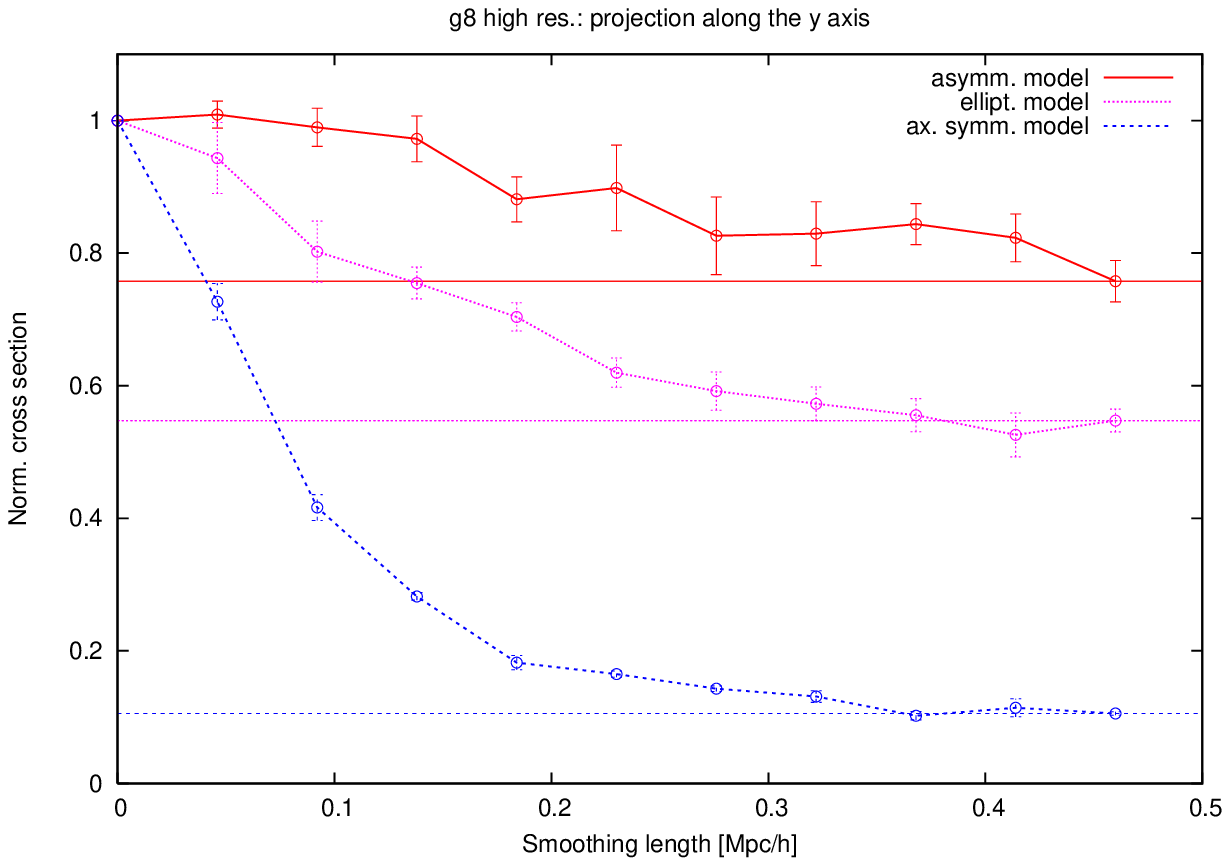}
  \includegraphics[width=0.33\hsize]{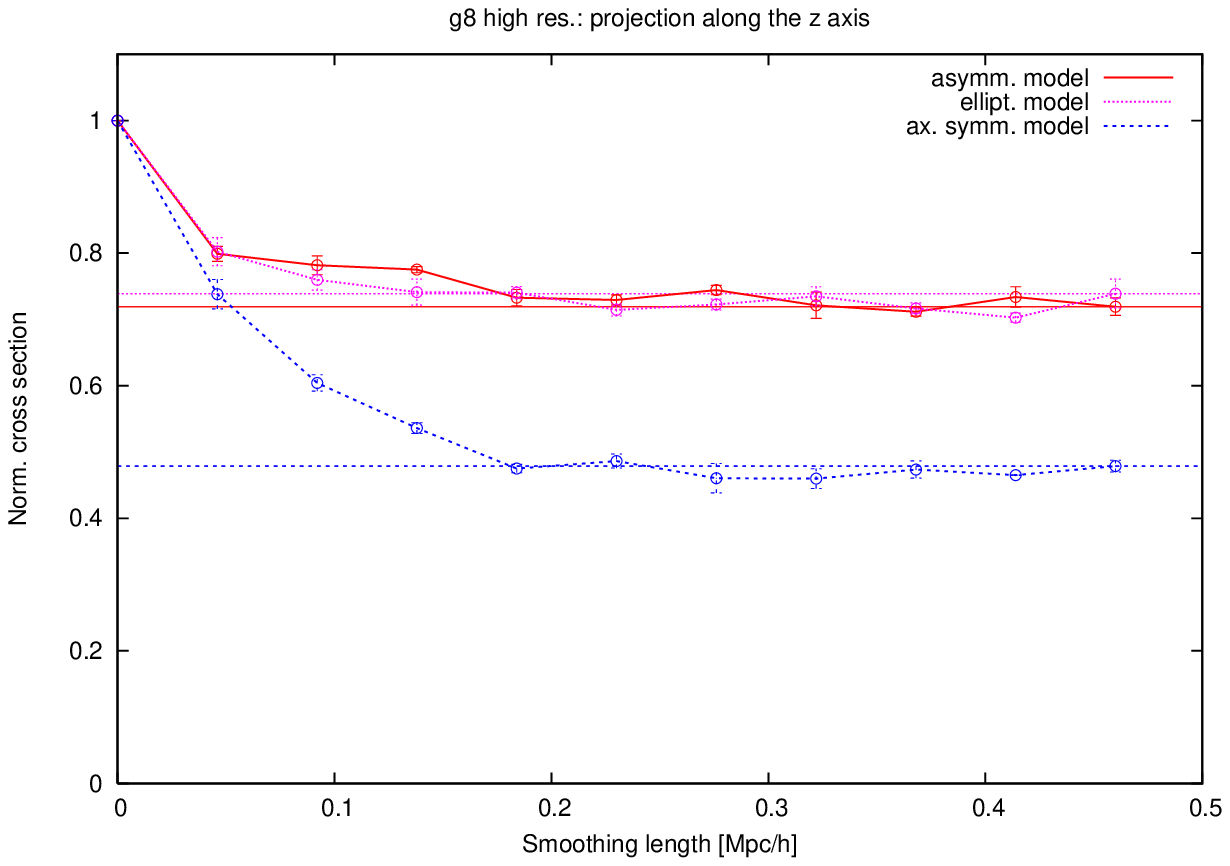}
\caption{Lensing cross section for arcs with length-to-width ratio larger than
  $7.5$ as a function of the smoothing length. Solid, dashed and dotted lines
  refer to smoothing adopting the asymmetric, the elliptical and the axially
  symmetric background models, respectively. The curves are normalised to the
  corresponding cross section for smoothing length equal to zero. Shown are
  the results for the cluster projections along the $x$- (left panel), $y$-
  (central panel), and $z$-axes (right panel). The critical lines
  in the three projections are contained in rectangles sized
  $(395\times 105)\,h^{-2}\mathrm{kpc^2}$, $(455\times
  205)\,h^{-2}\mathrm{kpc^2}$ and $(262\times
  87)\,h^{-2}\mathrm{kpc^2}$, respectively.}
\label{fig:avcs_smooth}
\end{figure*} 

\subsection{Lensing cross sections}

Aiming at quantifying the differences between the strong-lensing efficiency of
clusters with different amounts of ellipticity, asymmetries and substructures,
we focus on the statistical distributions of the arc length-to-width
ratios. Indeed, the distortion of the images of background galaxies lensed by
foreground clusters is commonly expressed in terms of these ratios. 

The efficiency of a galaxy cluster for producing arcs with a given property
can be quantified by means of its lensing cross section. This is the area on
the source plane where a source must be placed in order to be imaged as an arc
with that property.

The lensing cross sections for large and thin arcs are computed as described
in detail in several previous papers \citep[see e.g]{ME05.1}. We consider here
the cross sections for arcs whose length-to-width ratio exceeds a threshold
$(L/W)_{\rm min}=7.5$, and refer to these arcs as {\em giant} arcs.

\subsubsection{A particular case: the cluster $g8_{\rm hr}$}

The impact of ellipticity, asymmetry and substructures on the lensing cross
section for giant arcs obviously depends on the particular projected mass
distribution of the lens. Large differences can be found even between
different projections of the same cluster. As an example, we show in
Fig.~\ref{fig:avcs_smooth} the lensing cross sections for giant arcs as a
function of the smoothing length for the three projections of cluster $g8_{\rm
hr}$. Results are shown for all the smoothing methods described earlier. The
cross sections have been normalised to that of the unsmoothed lens. The
horizontal lines in each plot indicate the limiting values reached when the
surface density maps are completely smoothed. Three different realizations of
background source distributions were used to calculate the errorbars.

As expected, the lensing cross sections decrease as the smoothing scale
increases. The decrement is generally rapid for small smoothing lengths, then
becomes shallower. 

The differences between the three projections are large. When smoothing with
an elliptical background model, maximal variations of the cross section of the
order of $40\%$ are found for the projections along the $x$- and 
$y$-axes. For these two projections, circularising the surface-mass
distributions reduces the cross section by $\sim
85-90\%$. However, while for the projection along the $x$-axis smoothing using
the asymmetric background model reduces the lensing cross
section by $35\%$, for the projection along the $y$-axis the
cross section becomes only $\sim 20 \%$ smaller. The differences between these
two projections can be explained as follows. First, as discussed earlier in
the paper, when projected along the $x$-axis, the cluster has important
substructures close to its centre. This is evident in Fig.~\ref{fig:powers}:
substructures are significant at radii between $\sim 400 - 800
\,h^{-1}$kpc.  On the other hand, when projected along the $y$-axis the
cluster exhibits significant substructures only at larger radii, $>1
\,h^{-1}$Mpc. Since strong lensing occurs in the very inner region of the
cluster, the impact of substructures close to the centre is larger than that
of substructures farther away. Second, in the projection along the $y$-axis,
the twist of the iso-density contours and the variations of their ellipticity
are significantly larger than for the projection along the $x$-axis (see
Fig.\ref{fig:ellphi}). Therefore, while for the projection along the $x$-axis
the deviation from a purely elliptical mass distribution is mostly due to the
effects of substructures, in the projection along the $y$-axis it is due to
both substructures ($\sim 20\%$) and asymmetries ($\sim 25\%$).
Asymmetries which are due to the presence of large-scale density
fluctuations distort the isodensity contours which are elongated in
some particular direction, varying their ellipticity and position
angle. Such large-scale modes contribute to the shear field of the
cluster, pushing the critical lines towards regions of lower surface
density and increasing their size. Consequently, the strong lensing
cross section also increases.

As shown in Sect.~\ref{sect:power}, when projected along the $z$-axis, the
cluster appears rounder. Consequently, a smooth axially symmetric
representation of this lens which conserves its surface density profile has a
lensing cross section for giant arcs which is only $50\%$ smaller than that of
the original cluster. Smoothing using the asymmetric or the elliptical
background models is equivalent and leads to a reduction of the lensing cross
section by $\sim 30\%$. The absence of significant differences between these
two smoothing schemes indicates that asymmetries play little role in this
projection, while the large substructure observed at $\sim
800\,h^{-1}$kpc from the centre has a significant impact on the lensing properties
of this lens, even being at relatively large distance from the region where
strong lensing occurs.

\begin{figure}[t]
  \includegraphics[width=\hsize]{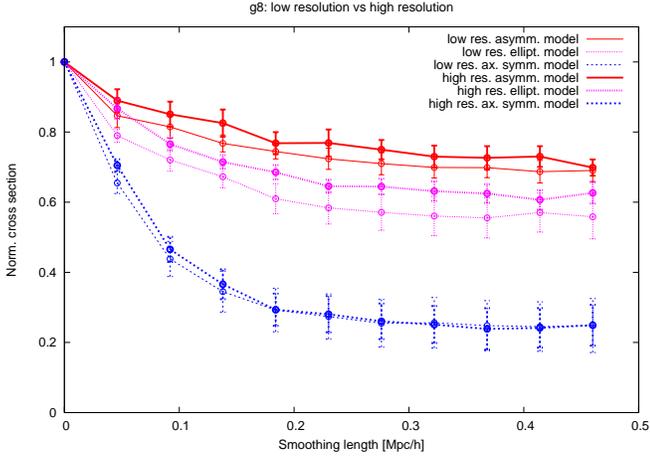}
\caption{Comparison between the low and the high resolution version of cluster
  g8. The lensing cross section for arcs with length-to-width ratio larger
  than $7.5$ averaged over three orthogonal projections of the same cluster
  are plotted versus the smoothing function.}
\label{fig:avcs_g8hl}
\end{figure} 

The smoothing length for which the curves converge to the values for the
completely smoothed maps tell us the characteristic scale of cluster
substructures which is important for lensing. In those projections where
localised substructures play an important role, i.e. in the projections along
the $x$- and the $z$-axes, the relevant scales are smaller ($\lesssim 100 -
300\,h^{-1}$kpc), while for the projection where asymmetries are more relevant
they are larger ($\lesssim 400\,h^{-1}$kpc). Converting these spatial scales
into the corresponding mass scales in not an easy task, especially because we
are dealing with substructures in two dimensions. Tentatively, we can assume
that the substructures are spherical and their mean density corresponds to the
virial overdensity $\Delta_v(z)$. For $z=0.3$, in the cosmological framework
where our simulations are carried out, $\Delta_v \sim 123$. Then, the above
mentioned spatial scales correspond to masses between $\sim 4\times 10^{11}$
and $\sim 2\times 10^{13} h^{-1}\,M_\odot$.

\begin{figure}[t]
  \includegraphics[width=\hsize]{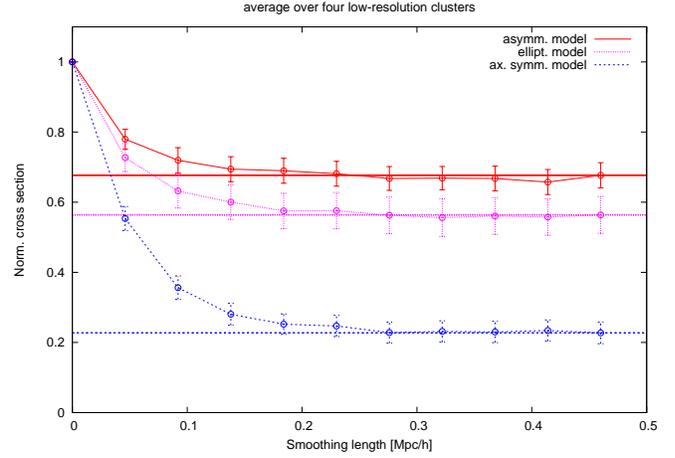}
\caption{Mean lensing cross section for arcs with length-to-width ratio larger than
  $7.5$ of four low-resolution clusters as a function of the smoothing length. Solid, dashed and dotted lines
  refer to smoothing adopting the asymmetric, the elliptical and the axially
  symmetric background models, respectively. The critical regions of the lenses have maximal
radii in the range $\sim 100 - 250 h^{-1}$kpc.}
\label{fig:avcs_low_all}
\end{figure} 

The three cluster projections whose lensing properties were
discussed above were carried out along the three orthogonal axes of
the simulation box. In general, these axes do not coincide with the
cluster's principal axes because it is randomly oriented with respect
to the simulation box. Thus, the roundest and the most elliptical
cluster projections that we have studied are not necessarily the
roundest and the most elliptical possible projections,
respectively. In the case of $g8_{hr}$, however, the principal axes do
not differ substantially from those of the simulation box. The cluster
turns out to be prolate with axis ratios $I_1/I_2 \sim 1.9$ and
$I_2/I_3 \sim 1.1$. When projected along the major principal axis,
i.e.~in its roundest projection, the ellipticity in the central region
is slightly smaller than in the projection along the $z$-axis, varying
between $0.1$ and $0.2$. When projected along the two other principal
axes, the cluster has ellipticity and twist profiles very similar to
those for the projections along the $x$- and the $y$-axes. For these
reasons, the differences between the strong lensing cross sections of
the purely elliptical and of the axially-symmetric smoothed models are
modest in the roundest projection, even smaller than for the
previously discussed projection along the $z$-axis. Indeed, we find
that the ellipticity accounts for only $10\%$ of the lensing cross
section in this case. When projected along the other two principal
axes, the impact of the ellipticity is similar to that for the
projections along the $x$- and $y$-axes.

\subsubsection{Mean lensing cross sections}

The example of cluster $g8_{\rm hr}$ shows that, depending on the particular
configuration of a lens, the impact of ellipticity, asymmetry and substructures
can be substantially different in different clusters. Nevertheless, we can try
to estimate what is the statistical impact of all these factors. For doing
that, we repeat the analysis shown for the cluster $g8_{\rm hr}$ on our sample
of clusters simulated with lower mass resolution. Among them, we analyse the
lensing properties of the low-resolution analogue of cluster $g8_{\rm hr}$. When
we compare the sensitivity to smoothing of the low- and the high-resolution
versions of the same cluster, we do not find significant differences between
them. Fig.~\ref{fig:avcs_g8hl} shows how the lensing cross section for giant
arcs changes as a function of the smoothing length for all three smoothing
schemes applied. Each curve is the average over the three independent
projections of the clusters. The thick and thin lines refer to the high-
and low-resolution runs, respectively. Considering that, as discussed in
Sect.~\ref{sect:nummod}, the two simulations have quite
different mass distributions, the differences shown here, which are still
within the error bars, have little significance. This suggests that
our results are not affected by problems of mass resolution of the numerical
simulations.

\begin{figure}[t]
  \includegraphics[width=\hsize]{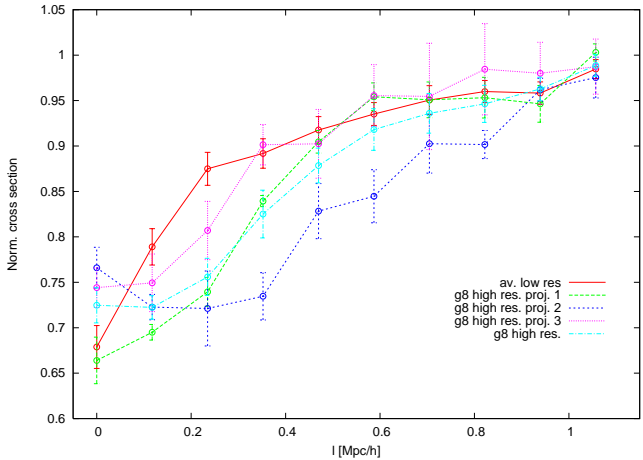}
\caption{Lensing cross section for arcs with length-to-width ratio larger than
  $7.5$ as a function of the minimal equivalent radius containing
  substructures. Results are shown for all three cluster projections. The
  curves are normalised to the cross section of the cluster containing all its
  substructures, corresponding to $l=\infty$. The critical regions of the
  lenses have maximal radii in the range $\sim 100 - 250 h^{-1}$kpc.}
\label{fig:avcs_sub}
\end{figure} 

In Fig.~\ref{fig:avcs_low_all} we show the variation of the lensing cross
section vs. smoothing length averaged over all the low-resolution clusters
which we have analysed. For each cluster, we use the three projections along
the $x$, $y$ and $z$ axes, i.e. 12 lens planes in total. The results are shown
for all three smoothing schemes adopted. The curves show that on average
removing the substructures from the clusters reduces their lensing cross
section by about $30\sim 35 \%$. Removing asymmetries, i.e. transforming the
cluster mass distributions to purely elliptical, further reduces the lensing
cross section for giant arcs by $\sim 10 \%$. If also ellipticity is removed,
the mean lensing cross section becomes $\sim 20\%$ of that of the unsmoothed
lenses. The typical scales for the substructures which mostly affect the
lensing properties of their host halos are $\lesssim 150\,h^{-1}$kpc ($M
\lesssim 10^{12}h^{-1}\,M_\odot$). Note that this does not mean that larger
substructures do not affect the lensing cross sections: simply, they are less
abundant. We verified that only one of the clusters in our sample ($g72$) is
undergoing a major merger with a massive substructure ($M_{\rm sub}\sim 4\times
10^{14}h^{-1}\,M_\odot$) at $z=0.3$. Note also
that the largest scale sub-halos contribute also to the asymmetry of the
projected mass distributions. This means that smoothing further using the
asymmetric background model does not remove these large substructures
completely. When smoothing using the background elliptical and
axially-symmetric models the smoothing length at which the lensing cross
section approximates that of the completely smoothed lenses is slightly
larger, because larger-scale contributions to the surface density fields must
be removed.

\subsubsection{Location of important substructures}

We now investigate what is the typical location of substructures which are
important for lensing. By ray-tracing through the mass distributions obtained
after removing substructures from outside a given equivalent radius, as
discussed at the end of Sect.\ref{sect:smooth}, we find that the strong
lensing efficiency of clusters is sensitive to substructures located within a
quite large region around the cluster centre. For demonstrating this, we show
in Fig.~\ref{fig:avcs_sub} how the lensing cross section changes as a function
of the minimal radius containing substructures. The cross sections are again
normalised to those of the unsmoothed lenses.

In the projection along the $y$-axis of cluster $g8_{\rm hr}$ (short dashed
line), we note that the lensing cross section decreases quickly when removing
substructures outside an equivalent radius of $\sim 1\,h^{-1}$Mpc. The lensing
cross section for giant arcs is already reduced by $\sim 10\%$ when the
minimal equivalent radius containing substructures is $\sim
800\,h^{-1}$kpc. In fact, in this projection there are two large substructures
at distances between $1$ and $1.2\,h^{-1}$Mpc from the cluster centre, which
seem to influence the strong lensing efficiency of this lens. Note that the
critical lines in this projection of the cluster extend up to $\sim 200
h^{-1}$kpc from the cluster centre. In the other two projections of the same
cluster (long dashed and dotted lines), where large substructures are located
closer to the centre, a similar decrement of the lensing cross section is
observed at much smaller equivalent radii, between $300$ and $450\,h^{-1}$kpc.

When averaging over the low-resolution cluster sample, we still find that the
lensing cross sections start to decrease when substructures outside a region
of equivalent radius $\sim 1\,h^{-1}$Mpc are removed from the clusters. While
the minimal radius containing substructures is further reduced, the cross
sections continue to become smaller. The evolution is initially shallow. A
reduction of $\sim 15\%$ is observed at a minimal equivalent radius $\sim
300\,h^{-1}$kpc. If substructures are removed from an even smaller region
around the centre of the clusters, the decrement of the lensing cross sections
becomes faster. The critical regions of the lenses in our sample have maximal
radii in the range $\sim 100 - 250 h^{-1}$kpc.

This shows that substructures close to the cluster centre are the most
relevant for strong lensing, but substructures located far away
from the cluster critical region for lensing also have a significant impact on
the cluster lensing cross sections. 

\subsection{Arc shapes, locations and fluxes}

\begin{figure}[t]
  \includegraphics[width=\hsize]{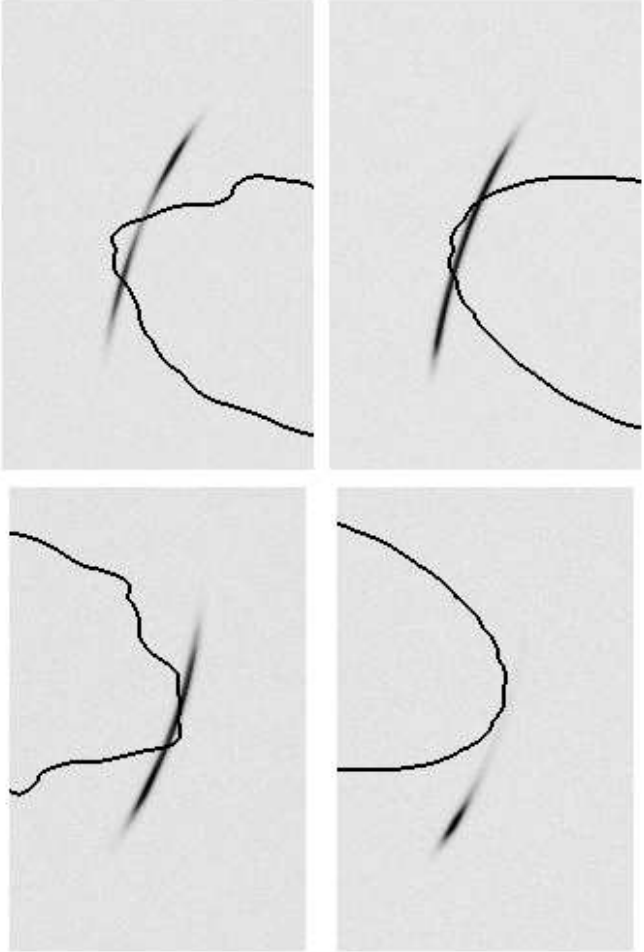}
\caption{Effects of substructures on scales $\lesssim 50\,h^{-1}$kpc on the
morphology of gravitational arcs. In the left panels shown are the arcs formed
out of two different sources lensed by the unsmoothed mass distribution
displayed in the top left panel of Fig.\ref{fig:incsmooth}. In the right
panels, shown are the corresponding images obtained by using as lens the mass
distribution given in the top right panel of Fig.\ref{fig:incsmooth}, which
has been obtained by smoothing with a smoothing length of $47\,h^{-1}$kpc. In
all panels, the critical lines are drawn for comparison. The top and
the bottom panels have sizes of $38''\times 57''$ and $36''\times 56''$,
respectively. }
\label{fig:splitjoinarcs}
\end{figure} 

Small changes in the positions of the caustics and therefore in the positions
of the critical lines can have huge consequences on the appearance and
location of gravitational arcs. In order to describe these effects we compare
here the characteristics of the images of the same population of background
sources lensed both with the unsmoothed projected mass distributions of the
numerical clusters and with weakly smoothed versions of them. We smooth using
the asymmetric model and using a smoothing length of $47\,h^{-1}$kpc ($M\sim
5\times 10^{10} h^{-1}\,M_\odot$), not significantly exceeding the scale of
galaxies in clusters. The aim of this discussion is to show that even
relatively small substructures may play a crucial role in determining the
appearance of gravitational arcs.

Sources are first distributed around the caustics of the unsmoothed lens
following the method discussed in Sect.~\ref{sect:raytr}. Then, the same
sources are used when ray-tracing through the surface density maps from which
substructures are removed. Each source conserves its position, luminosity,
ellipticity and orientation, allowing to directly measure the effects that
removing substructures and asymmetries has on several properties of the same
arcs. For this experiment, we use an extended version of our ray-tracing code
which includes several observational effects, like sky brightness and photon
noise, allowing to mimic observations in several photometric bands. We assume
that the sources have exponential luminosity profiles and shine with a
luminosity in the $B$-band $L_B=10^{10}\,L_{\odot}$. We simulate exposures of
$3$ksec with a telescope with diameter of $8.2$m (VLT-like). The throughput of
the telescope has been assumed to be $0.25$. The surface brightness of the sky
in the $B$-band has been fixed at $22.7$mag per square arcsec. In this ideal
situation, no seeing is simulated. The effects of all these observational
effects on the morphological properties of arcs will be discussed in a
forthcoming paper.

\begin{figure}[t]
  \includegraphics[width=\hsize]{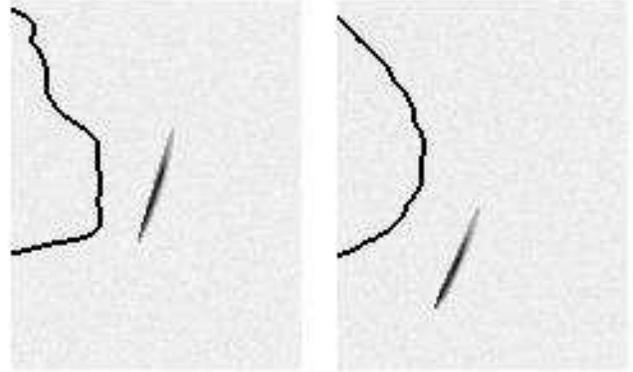}
\caption{Example of gravitational arc shifted by substructures. The size of
the each frame is $27''\times 34''$. Left panel: simulation including
substructures. Right panel: simulation performed after smoothing with a
smoothing length of $47\,h^{-1}$kpc.}
\label{fig:arcshift}
\end{figure}

In earlier studies, \cite{ME00.1} and \cite{FL00.1} showed that the impact of
galaxy-sized cluster subhalos on the statistical properties of gravitational
arcs with large length-to-width ratios is very modest. These results are
confirmed in the present study. As shown in the previous section, the lensing
cross sections for long and thin arcs decrease by $\sim 20\%$ when smoothing
the cluster surface densities on scales $\sim 50\,h^{-1}$kpc. On smaller
scales the decrement is only of a few percent. However, the morphology of
individual arcs is strongly affected in several cases. Arcs can become longer
or shorter, thinner or thicker. In other cases, more dramatic morphological
changes are found. For example, cluster galaxies locally perturb the cluster
potential such as to break long arcs, while in other cases the
opposite effect occurs. Two examples are shown in
Fig.\ref{fig:splitjoinarcs}. In the upper panels, the same source is imaged as
two short arcs or as a single long arc when the unsmoothed (left panel) and
the smoothed lens (right panel) are used, respectively. The lenses displayed
in the top panels of Fig.~\ref{fig:incsmooth} have been used for these
simulations. In the bottom panels, a single long arc becomes a smaller arclet
in absence of substructures. Super-imposed on each graph are the critical
lines. They tend to wiggle around individual substructures in the left panels,
while they are more regular in the right panels. Substructures slightly shift
the high-magnification regions of a cluster relative to the background
sources, inducing remarkable changes in the shape of their images and in their
multiplicity. For the cluster projection used in this example, the image
multiplicity is increased for $\sim 21\%$ of the sources producing
arcs longer than $5''$, when smoothing is applied, indicating that long arcs
break up. On the other hand, for $\sim 10\%$ of them the image multiplicity
decreases, showing that the caustics shrink and sources move outside of them,
Consequently the number of their images decreases.

In several cases, substructures are also responsible for significant shifts of
the positions of gravitational arcs. An example is shown in
Fig.~\ref{fig:arcshift}. The size of each frame is $27''\times 34''$. The
morphological properties of the arc in the two simulations are almost
identical. The arc length is $\sim 11''$, the arc width is $\sim0.6''$. The
luminosity peak of the arc, which we use for measuring the shift, is moved
towards the bottom left corner of the frame by $\sim 8.5''$, when substructures
on scales smaller than $47\,h^{-1}$kpc are smoothed away. Similar cases are
frequent. For the lens used in this example, $\sim 27\%$ of the long arcs
(length $>5''$) found in the simulation including substructures are shifted by
more than $5''$ after smoothing. About $\sim 4\%$ of
them is shifted by more than $10''$. From this analysis, long arcs which split
into smaller arclets are excluded.

\begin{figure}[t]
  \includegraphics[width=\hsize]{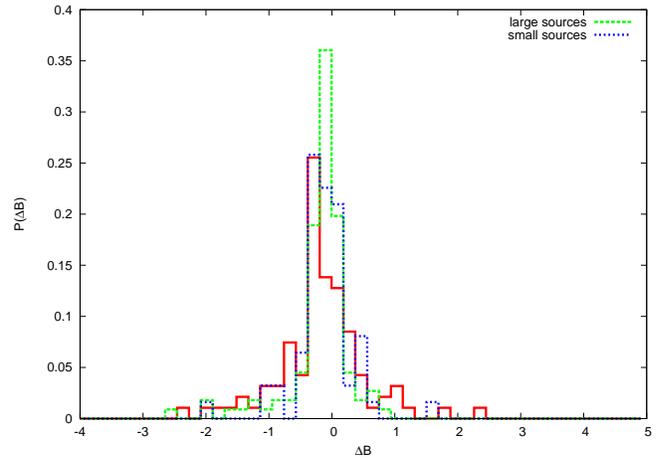}
\caption{Probability distribution function of the differences between arc
magnitudes in simulation including and excluding substructures on scales
$<47\,h^{-1}$kpc. Results are shown for three different source sizes. See text
for more details.}
\label{fig:fluxhist}
\end{figure}

Finally, substructures affect the fluxes received from the lensed sources. The
histogram in Fig.~\ref{fig:fluxhist} (solid line) shows the probability
distribution function of the differences $\Delta B=B-B_{sm}$ between
magnitudes of arcs with length $>5''$ measured in the simulations where the
unsmoothed and smoothed lens projected mass maps were used as lens planes,
respectively. The analysis is restricted to arcs whose length exceeds $5''$ in
the simulation containing all substructures. Some arcs are magnified, some
others demagnified by the substructures. The maximal variations in luminosity
correspond to $\Delta B \sim -2.3 \div +2.4$. The distribution is slightly
skewed towards the negative values, indicating that in absence of
substructures arcs tend to be less luminous. In fact, substructures contribute
to magnify the sources, as discussed in Sect.~\ref{sect:mag}. 

Sources of different size are expected to be differently
susceptible to the substructures. The dashed and the dotted lines in
Fig.~\ref{fig:fluxhist} respectively show how the probability
distribution function of $\Delta B$ changes when the source size is
increased or decreased by a factor of two compared to the original
source size used in the simulations. As expected, larger sources are
less sensitive to perturbations by small substructures in the lenses.

Similar results were found for some other cluster models. For other lenses,
the impact of the substructures on the properties of individual arcs is even
stronger. 

\begin{figure}[t]
  \includegraphics[width=\hsize]{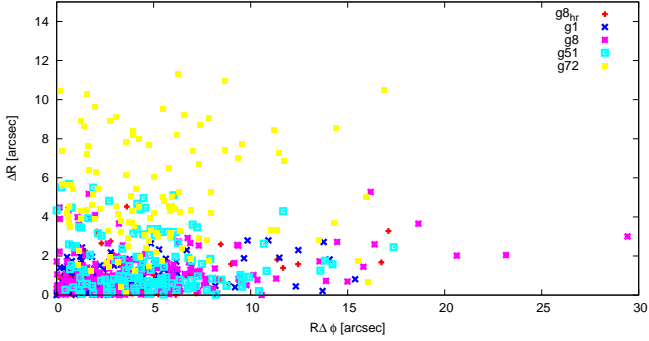}
\caption{Distribution of long arcs (length $>5''$) in the plane radial
  ($\Delta R$) vs.~tangential ($R\Delta \phi$) shift. Different
  symbols identify arcs produced by different clusters. The arcs
  produced by the cluster $g72$, which is experiencing a major merger,
  are given by the small filled squares.}
\label{fig:radtanshifts}
\end{figure}

The observed arc shifts have tangential and radial
components. Generally, the tangential shifts are larger than the
radial shifts. However, when large substructures located close to the
critical regions of clusters are smoothed away, significant radial
shifts are possible, given that the relative size of the critical
lines changes dramatically. In Fig.~\ref{fig:radtanshifts}, the radial
shifts of long arcs (length $>5''$) is plotted versus the tangential
shifts. Different symbols are used to identify arcs produced by
different numerical clusters. As anticipated, for the majority of the
arcs produced by the clusters $g1$, $g8$, $g8_{hr}$ and $g51$ the
radial shifts are within few arcseconds, while tangential shifts of
$10''$ and more are frequent. On the other hand, the arcs produced by
the cluster $g72$ have significantly larger radial shifts. As
mentioned above, $g72$ is undergoing a major merger and a secondary
lump of matter occurs near the cluster centre. The cluster critical
lines, along which arcs form, are elongated towards it. When moderate
smoothing is applied, the impact of the merging substructure is
attenuated and the critical line shrinks substantially. Thus, the arcs
move towards the centre of the cluster and their morphology and flux
are also strongly affected.

\begin{figure}[t]
  \includegraphics[width=\hsize]{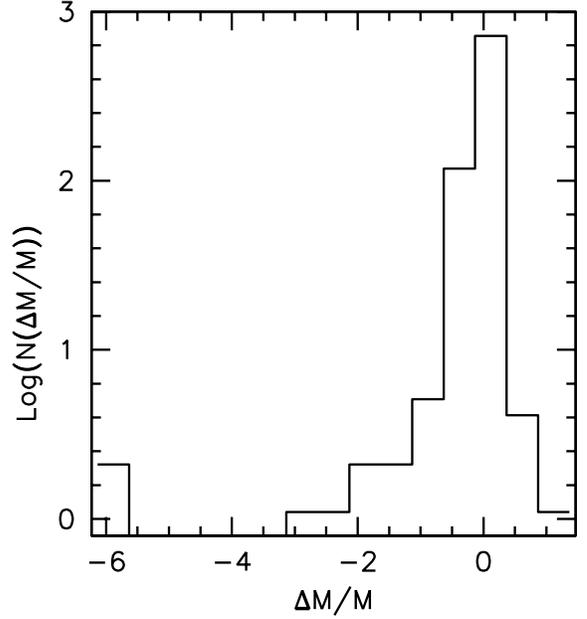}
\caption{Distribution of the relative variations of mass determinations from
  strong lensing, assuming axial symmetry and that the arc position trace the
  location of the critical lines.}
\label{fig:mbias}
\end{figure}

Some results for all the cluster models we analysed are summarised in
Tab.\ref{tab:subarcs}. All of these effects might have an enormous impact in
lensing analysis of clusters, in particular when modelling a lens by fitting
gravitational arcs. These results show that any substructure on scales
comparable to those of galaxies should be included in the model in order to
avoid systematic errors. This problem will be addressed in detail in a
following paper, in particular regarding the possible biases in strong
lensing mass determinations. However, by making the wrong assumption of axial
symmetry, we can approximately estimate the errors due to the radial shifts
of the arcs. For axially symmetric lenses, the mean convergence within the
critical line is $\overline\kappa(<x_c)=1$. The mass within $x_c$
 is then
\begin{equation}
  M(<x_c)=\pi \Sigma_{\rm cr} x_c^2 \;,
\end{equation}
where 
\begin{equation}
\Sigma_{\rm cr}= \frac{c^2}{4 \pi G}
      \frac{D_{\rm s}}{D_{\rm l} D_{\rm ls}}
\end{equation}
is the critical surface mass density and $x_c$ is in physical units. We assume
that the position of an arc traces the position of the critical line. Then, if
an arc distance from the centre changes from $R$ to $R'$, the relative
variation of the mass inferred from strong lensing is 
\begin{equation}
  \frac{\Delta
  M}{M}=\frac{M-M'}{M}=\frac{R^2-R'^2}{R^2}\;. 
\end{equation}
The distribution of such $\Delta M/M$, as derived from the radial shifts
displayed in Fig.\ref{fig:radtanshifts}, is shown in Fig.~\ref{fig:mbias}. 
Without suitably modelling the effects of substructures, the typical
errors in mass determinations are within a factor of two, but larger
errors are also possible. Since substructures generally contribute to enlarge
the critical lines, a larger mass within the critical line would be required
in order to have an arc at the observed distance from the cluster
centre.   

Note that even substructures far away from the cluster centre are
important. For example, keeping the inner structure of the projection along
the $y$-axis of cluster $g8_{\rm hr}$ unchanged, while removing the big
substructures at distances $> 1\,h^{-1}$Mpc, we find that more than $\sim
50\%$ of the long arcs are shifted by at least $5''$. Moreover, image
multiplicity increases for $26\%$ and decreases for $8\%$ of the sources
respectively.       

\begin{table}[hbtp]
\centering
\caption{Effects of substructures on gravitational arcs. Column 1: cluster
name; column 2: projection; column 3: percentage of sources whose image
multiplicity increases; column 4: percentage of sources whose image
multiplicity decreases; column 5: percentage of long arcs ($l>5''$), whose
positions result to be shifted by more than $5''$ when substructures are
smoothed away; column 6: maximal variations of magnitudes of long arcs.}
\begin{tabular}{cccccc}
\hline \hline
Cluster      & proj. & inc. mult.  & dec. mult. & shift $>5''$ & $\Delta B$ \\
             &       & [$\%$]      &  [$\%$]    &  [$\%$]      &            \\
\hline \hline
$g8_{hr}$    & x     &  21.2       &  9.8       &  26.6        &  -2.3/+2.4 \\
             & y     &  19.7       &  3.9       &  11.1        &  -1.7/+0.8 \\ 
             & z     &  23.7       &  10.1      &  15.5        &  -2.5/+1.6 \\
\hline
$g1$         & x     &  47.0       &  1.0       &  24.0        &  -1.9/+1.1 \\
             & y     &  26.1       &  0.0       &  20.0        &  -1.7/+0.7 \\ 
             & z     &  24.5       &  0.0       &  29.5        &  -2.5/+0.8 \\
\hline
$g8$         & x     &  24.0       &  4.0       &  27.0        &  -1.7/+1.0 \\
             & y     &  35.4       &  9.5       &  51.1        &  -2.1/+2.1 \\ 
             & z     &  31.2       &  6.1       &  28.2        &  -2.5/+0.5 \\
\hline
$g51$        & x     &  33.1       &  6.2       &  44.7        &  -2.3/+1.7 \\
             & y     &  35.6       &  5.1       &  65.7        &  -2.5/+1.3 \\ 
             & z     &  39.1       &  6.2       &  54.3        &  -1.5/+1.3 \\
\hline
$g72$        & x     &  36.0       &  4.0       &  79.4        &  -2.5/+2.1 \\
             & y     &  62.5       &  0.0       &  57.1        &  -2.4/+0.0 \\ 
             & z     &  22.5       &  0.0       &  70.9        &  -2.4/+1.7 \\
\hline
mean         &       &  32.1       &  4.4       &  40.3        &  -2.1/+1.3 \\
\hline \hline
\end{tabular}
\label{tab:subarcs}
\end{table}

If relatively small substructures can alter many of the properties of
gravitational arcs, even asymmetries may be relevant. As noted earlier the
projection along the $y$-axis is the most asymmetric of $g8_{hr}$. Comparing
the properties of arcs lensed by the smoothed asymmetric and elliptical models
of this lens, we find significant shifts in the location of about $45\%$ of
the long arcs. For $\sim 20\%$ of the sources producing long arcs, the
multiplicity is changed.
  
\section{Conclusions}
\label{sect:conclu}

In this paper we have quantified the impact of several properties of
realistic cluster lenses on their strong lensing ability. In particular, our
goal was to separate the effects of substructures, asymmetries and ellipticity.
For doing that, we analysed the lensing properties of one numerical cluster
simulated with very high mass resolution. In addition, we studied four other
clusters obtained from N-body simulation with a lower mass resolution. 

Each cluster was projected along three independent directions.  For each
projection, we constructed three completely smoothed versions. Each of them
conserves the mean surface density profile of the mass distribution of the
cluster. However, the first reproduces the variations of the ellipticity and
of the position angle of the isodensity contours as functions of the distance
from the centre; the second has elliptical isodensity contours with fixed
ellipticity and orientation; the third is an axially symmetric model.

The lensing properties of the numerical clusters, of their smoothed analogues
and of several intermediate versions were investigated using standard
ray-tracing techniques.

Our main results can be summarised as follows:
\begin{itemize}
\item Substructures, asymmetries and ellipticity contribute to increase
the ability of clusters to produce strong lensing events. Substructured,
asymmetric and highly elliptical clusters produce more extended high
magnification regions in the lens plane where long and thin arcs can
form. Indeed, substructures, asymmetries and ellipticity determine the
location and the shape of the lens caustics around which sources must be
located in order to be strongly lensed by the clusters.
\item The impact of substructures, asymmetries and ellipticity on the lensing
cross section for producing giant arcs is different for different lenses. The
lensing properties of the most symmetric clusters appear to be particularly
influenced by the substructures. On the contrary, substructures are less
important in asymmetric lenses. 
\item On average, we quantify that substructures
account for $\sim 30\%$ of the total cluster cross section, asymmetries for
$\sim 10\%$ and ellipticity for $\sim 40\%$.
\item The substructures that typically contribute to lensing are on scales
$\lesssim 150 - 200 \,h^{-1}$kpc. Assuming a virial overdensity of $\sim
123$ for $z=0.3$, this corresponds to mass scales of the order of $\sim
10^{12} h^{-1}\,M_\odot$. Substructures on larger scales are not as frequent
in our cluster sample, but, if present, they can boost significantly the
lensing cross section \citep[see e.g.][]{TO04.1,ME05.1}.
\item Substructures play a more important role when they are located close
to the cluster centre. However, the lensing cross section for giant arcs is
sensitive to substructures within a wide region around the cluster core. In
particular, our simulations show that the sensitivity to substructures far
from the centre is particularly high in those clusters whose inner regions are
unperturbed. In these cases, the loss of strong lensing efficiency due to
removing the substructures from the clusters is correlated with substructures
within a region of $\sim 1 \, h^{-1}$Mpc in radius; on the contrary, clusters
containing substructures in the inner regions are ``screened'' against
external perturbers.
\item Even small substructures ($l\lesssim 50$kpc, $M\lesssim 5\times
10^{10}h^{-1}\,M_\odot$) influence the appearance and the
location of gravitational arcs. The perturbations to the projected
gravitational potential of the cluster induced by the substructures
alter the multiplicity of the images of individual sources. Moreover, they
change the morphology and the flux of the images themselves. Finally, they can
shift the position of arcs with significant length to width ratios by several
arcseconds on the sky.    
\end{itemize}  

These results highlight several important aspects of strong lensing by
clusters. First, any model for cluster lenses cannot neglect the effects
of asymmetries, ellipticity and substructures. Clusters which may appear as
relaxed and symmetric, for example in the X-rays, are potentially those
which are most sensitive to the smallest substructures, located even at large
distances from the inner cluster regions, critical for strong lensing. Even
subhalos on the scales of galaxies can influence the strong lensing
properties of their hosts and alter the shape and the fluxes of gravitational
arcs. Therefore, if the lens modelling is not carried out at a very high level
of detail, it may result in being totally incorrect.

Second, the high sensitivity of gravitational arcs to deviations from regular,
smooth and symmetric mass distributions suggests that strong gravitational
lensing is potentially a powerful tool to measure the level of substructures
and asymmetries in clusters. Since, as we said, the sensitivity to
substructures is higher in the case of more symmetric lenses, we conclude that
dynamically active clusters, like those undergoing major merger events, should
be quite insensitive to ``corrugations'' in the projected mass distribution
but highly sensitive to asymmetries. Arcs could then be used to diagnose
mergers in clusters. Conversely, substructures should become increasingly
important for the arc morphology as clusters relax. Then the level of
substructures in clusters should be quantified by measuring their effect on
the arc morphology. This is particularly intriguing since measuring the
fine structures of gravitational arcs has become feasible thanks to the high
spatial resolution reached in observations from space.

Third, the strong impact of asymmetries and substructures on the lensing
properties of clusters and the wide region in the cluster where these last can
be located in order to produce a significant effect further support the
picture that mergers might have a huge impact on the cluster optical depth for
strong lensing, as suggested in several previous studies
\citep{TO04.1,ME04.1,FE05.1}. 

\acknowledgements{We are grateful to the anonymous referee for his helpful
  comments and suggestions. The $N$-body simulations were performed at the
  ``Centro Interuniversitario del Nord-Est per il Calcolo Elettronico''
  (CINECA, Bologna), with CPU time assigned under an INAF-CINECA grant. This
  work has been supported by the Vigoni programme of the German Academic
  Exchange Service (DAAD) and Conference of Italian University Rectors
  (CRUI). F.P. is supported by the German Science Foundation under grant
  number BA 1369/5-2.}

\bibliography{../TeXMacro/master}
\bibliographystyle{aa}

\end{document}